\documentclass[preprint,12pt]{elsarticle}

\usepackage{color}
\usepackage{multirow}
\usepackage{balance}
\usepackage{hyperref}
\usepackage{siunitx}
\usepackage{amssymb,amsmath}
\usepackage{enumitem}
\usepackage{gensymb}
\usepackage{graphicx}
\usepackage{pdfsync}
\usepackage{algorithm}
\usepackage{algorithmicx}
\usepackage{algpseudocode}

\usepackage{graphicx}  

\newcommand\textBlack{\textcolor{black}}
\usepackage{booktabs}
\usepackage{breakurl}
\usepackage{multirow}
\usepackage[table,x11names]{xcolor}

\usepackage{adjustbox}
\usepackage{cleveref}

\journal{Applied Energy}

\begin{document}
\begin{frontmatter}

\title{A Valuation Framework for Customers Impacted by Extreme Temperature-Related Outages}



\tnotetext[label1]{This work is supported by the U.S. Department of Energy (DOE), Office of Electricity, Grid Controls and Communications Program. Pacific Northwest National Laboratory, operated by Battelle for the U.S.DOE under Contract DE-AC05-76RL01830.}

\author[inst1]{Min Gyung Yu} 
\author[inst1]{Monish Mukherjee}
\author[inst1]{Shiva Poudel}
\author[inst1]{Sadie R. Bender}

\affiliation[inst1]{organization={Pacific Northwest National Laboratory},
            city={Richland},
            postcode={99352}, 
            state={WA},
            country={USA}}
            
\author[inst2]{Sarmad Hanif,}

\affiliation[inst2]{organization={Electric Power Research Institute},
            city={Washington, DC},
            postcode={20005}, 
            state={},
            country={USA}}
            
\author[inst1]{Trevor D. Hardy}
\author[inst1]{Hayden M. Reeve}


\begin{abstract}
\textcolor{black}{Extreme temperature outages can lead to not just economic losses but also various non-energy impacts (NEI), such as increased mortality rates, property damage, and reduced productivity, due to significant degradation of indoor operating conditions caused by service disruptions. However, existing resilience assessment approaches lack specificity for extreme temperature conditions. They often overlook temperature-related mortality and neglect the customer characteristics and grid response in the calculation, despite the significant influence of these factors on NEI-related economic losses. This paper aims to address these gaps by introducing a comprehensive framework to estimate the impact of resilience enhancement not only on the direct economic losses incurred by customers but also on potential NEI, including mortality and the value of statistical life during extreme temperature-related outages. The proposed resilience valuation integrates customer characteristics and grid response variables based on a scalable grid simulation environment. This study adopts a holistic approach to quantify customer-oriented economic impacts, utilizing probabilistic loss scenarios that incorporate health-related factors and damage/loss models as a function of exposure for valuation. The proposed methodology is demonstrated through comparative resilient outage planning, using grid response models emulating a Texas weather zone during the 2021 winter storm Uri. \textBlack{The case study results show that enhanced outage planning with hardened infrastructure can improve the system resilience and thereby reduce the relative risk of mortality by 16\% and save the total costs related to non-energy impacts by 74\%.} These findings underscore the efficacy of the framework by assessing the financial implications of each case, providing valuable insights for decision-makers and stakeholders involved in extreme-weather related resilience planning for risk management and mitigation strategies.}
\end{abstract}
\begin{keyword}
Resilience, extreme temperatures, power distribution control, valuation, customer interruption costs
\vspace{-0.2cm}
\end{keyword}

\end{frontmatter}

\section{Introduction}
In recent years, there has been a notable increase in weather-related power outages driven by the impacts of climate change. The average annual number of such outages increased by approximately 78\% between 2011-2021 compared to 2000-2010 \cite{ClimateCentral}. 
\textcolor{black}{Of the major power outages since 2000, an overwhelming majority of them were due to weather-related events \cite{ccentral}. Such, outages caused by severe weather events have caused widespread damage to power infrastructure and affected a significant number of customers \cite{Abi2011}. While all outages can be costly, the concurrence of extreme weather introduces additional risks and costs for all actors within the system.}  In 2021, the extreme winter storm event in Texas resulted in a loss of power for more than 4.5 million homes and recent analysis links nearly 200 deaths \& \$18 billion in property damage to the blackout \cite{FED}. The grid failures associated with a heat wave in the Pacific Northwest of the U.S. during June 2021, resulted in a loss of power to tens of thousands of customers, at least 600 excess deaths, and more than 3,500 emergency visits for heat-stress related-illness \cite{NW_heat_wave}. As the frequency and intensity of extreme temperature events are expected to increase 
\cite{dahl2019increased}, the policymakers and utilities need methodologies and data for a holistic approach to estimate the true value of system resilience towards supporting policy and investment decision-making that can effectively improve resilience.

Extensive research efforts have been focused on the planning, operation, and valuation of power system resilience 
in response to extreme events (e.g., hurricanes, earthquakes, and floods), and various man-made attacks (cyber and physical threats) \cite{Ren2021, 8966351}. However, planning and evaluating the resilience of power systems in the context of extreme temperature-related events such as heatwaves and cold waves poses a unique and specific set of challenges. These challenges stem from the compounding effects of temperature extremes, which can result in increased risks associated with mortality and other severe consequences (e.g., property damage and medical expenses) \cite{morss2011}. 

In related literature, several methods have been proposed on the resilience valuation (in monetary value) of customer interruption to express the disruption incurred due to a service outage. \textcolor{black}{Many system improvements are justified by determining the value of lost load (VOLL) which represents the costs of interrupted electrical supply or willingness to pay to avoid an interruption, typically measured in per kWh \cite{schroder2015value}. While VOLL is appropriate for estimating capacity values at peak times, it is not generally proper for use at the distribution level, where interruptions occur with random frequency over various times of day and year \cite{wakefield2012guidebook}. To facilitate utilities with planning and operation for resilience, some research efforts focused on utilizing survey-based cost data to develop the damage functions relating to interruption cost based on outage duration and customer type \cite{Wacker1989}.} To further enhance cost metrics, studies were conducted to differentiate individual customer damage functions based on customer sectors (e.g., commercial, industrial, residential) and income levels to determine an optimal and reliable level of customer service \cite{Ali1999, Billinton2005}. 
One notable tool in this regard is the Interruption Cost Estimator (ICE) Calculator \cite{ICE, ICEdoc}, which assists in estimating the potential financial impact.
However, it is important to note that the cost models used in the ICE calculation have limitations in their scope, and they are not suitable for estimating costs for outages lasting longer than 16 hours \cite{ICEdoc}. \textcolor{black}{
While these methods can be used to estimate the customer economic losses with reasonable accuracy, they overlook the impact of partial/intermittent service, the vulnerability of communities, and many of the costs and non-energy impacts incurred by customers. Hence, using them as a singular criterion for resilience planning decisions may lead to inequitable social outcomes \cite{gorman2022quest}.}

\textcolor{black}{During extreme temperature events, a service outage may lead to additional risks due to increased mortality rates or property damage associated with loss of power (e.g., burst pipes and foundation damage) \cite{zuzak2022national}}. For example, individual houses may lose heating completely if they require electricity, which could result in indoor temperatures rapidly dropping and uncomfortable thermal conditions for occupants. In many households, electricity plays a crucial role in heating systems. Even traditional natural gas furnaces often rely on electricity to distribute the heated air, and some fireplaces are equipped with fans to enhance heat distribution within a room. Virtually all houses, whether using electric heating, gas heating, or even traditional fireplaces, are reliant on electricity to ensure efficient and effective heating. As the United States continues its decarbonization efforts, an increasing number of houses are likely to adopt electric heating systems, emphasizing the importance of electricity in maintaining comfortable indoor temperatures during extreme temperature conditions.
Additionally, buildings with insufficient insulation in walls, ceilings, and floors are more vulnerable to the effects of extreme cold weather, due to heat loss through the building \cite{BTOPNNL}. Exposure to extreme cold temperatures can lead to several health problems, including frostbite, hypothermia, respiratory problems, heart problems, and infections (e.g., flu, cold). During the 2021 outage in Texas due to Winter Storm Uri, hypothermia and frostbite accounted for 65\% of deaths, as people died of hypothermia inside their homes without power \cite{TexasNews}. This is particularly relevant for economically vulnerable populations, who may have less access to healthcare resources, proper heating, and well-insulated housing, making them vulnerable to extreme temperature impacts. Additionally, occupants with pre-existing medical conditions within these communities are at even greater risk, as the cold can exacerbate their health issues and may lead to more severe complications. 
Research has shown that severe temperature changes can have a significant impact on mortality rates, and prolonged exposure to extreme temperatures can increase the risk of temperature-induced illnesses \cite{McGeehin2001}. \textcolor{black}{Therefore, considering customer safety and social consequences along with the economic losses is essential towards a more complete evaluation of the benefits associated with improved resilience \cite{REG}.} 

\textcolor{black}{Existing research has made some progress in the area of evaluating the health risks of customers during compound climate and infrastructure failure events. The work in \cite{gasparrini2015mortality} identified the effects of mild and extreme weather through temperature–mortality associations that were conducted for 384 locations and presented the importance of considering public health implications. The study in \cite{ stone2023blackouts} evaluated the public health impacts during extreme heat waves using heat-related mortality simulated based on EnergyPlus building models.  Some existing work also assessed the thermal resilience of buildings through metrics, like survivability and hours of habitability, during power outage events \cite{BTOPNNL} and demonstrated the impact of better insulation and backup power in improving resilience \cite{sheng2023assessing}. }
\textcolor{black}{Although these studies have made some progress in the area of assessing customer safety during extreme weather events, prior work falls short in three crucial aspects: (1) the evaluation process does not provide a direct pathway for translating electric grid performance into social and economic consequences (2) the modeling frameworks do not incorporate the customer characteristics (e.g.,  insulation-level, fuel sources) and grid response variables  (e.g., physics-based load models, weather-conditions, energy availability, system capacity constraints, etc.) 
and (3) the approaches lack scalability and specificity to evaluate the non-energy impacts of customers in a specific region during extreme 
conditions.} These limitations can be significant as it is challenging to generate ``what-if" scenarios necessary for grid resilience planning and investment justifications.

This work proposes a comprehensive valuation framework for customer resilience that incorporates the non-energy impacts including mortality and value of statistical life (VSL) along with direct energy-related impacts (e.g. interruption costs, etc.) for customers. The proposed framework is complimented with scalable grid-response models to demonstrate how the framework can utilized towards evaluating the impact of resilience enhancement strategies. The key contribution of this work is summarized as follows: 
\begin{itemize}
    \item Resilience valuation framework designed specifically for outages caused by extreme temperature events. 
    \item Incorporating non-energy impacts like mortality as a significant consequence and cost in customer valuation during extreme temperature outages. 
    \item Comparative demonstration study for resilient outage planning by quantifying comprehensive costs of power outages by utilizing an at-scale simulation, emulating Texas freeze event during winter storm Uri, to represent resilience scenarios.
\end{itemize}

\section{Customer Valuation Framework for Extreme Temperature-Related Outage}
\label{sec:value_models}

\begin{figure}[!t]
\centering
\makebox[\textwidth][c]{\includegraphics[width=1.1\textwidth]{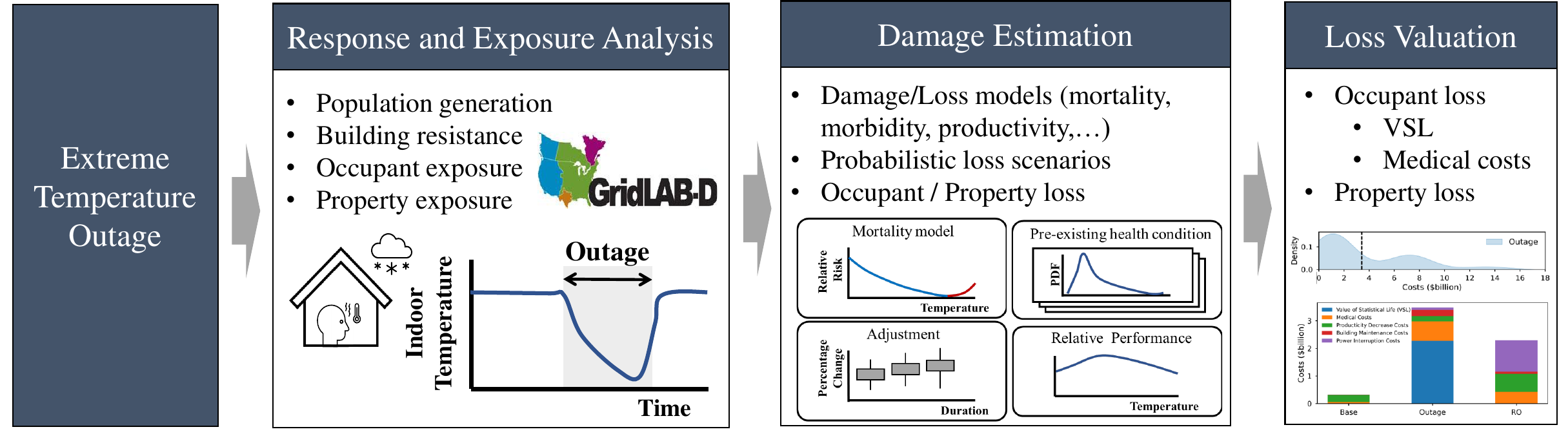}}
 \vspace{-0.5 cm}
\caption{Overview of customer interruption cost valuation}
\label{fig:overview}
 \vspace{-0.5 cm}
\end{figure}

\subsection{Overview}
This study proposes a framework to estimate the value of customer interruption due to extreme temperature-related outages. The overall procedure for estimating the related damages and losses can be divided into four stages, where the outputs of each stage serve as inputs for the next stage, as shown in Fig. \ref{fig:overview}. All stages are briefly explained below:
\begin{itemize}[leftmargin=0.35 cm]
    \item ``Extreme temperature outage" defines the weather and outage data necessary, typically using historical values. 
    \item ``Response and exposure analysis" generates the building populations for simulation and analyzes the exposure of each customer during extreme temperatures. 
    This study uses the Transactive Energy Simulation Platform (TESP), a co-simulation platform that facilitates emulating at-scale distribution grid response 
    \cite{TESP}. With physics-based modeling, the response model enables emulating the indoor conditions of individual houses considering their characteristics under various weather and grid conditions. By utilizing this approach, we can analyze the performance of buildings and determine whether and to what extent individuals are exposed to extreme temperatures inside the building. 
    \item ``Damage estimation" assesses the vulnerabilities of populations. Simulation studies provide input for estimating the associated impacts and damages to buildings and occupants. Detailed function models associated with damages, such as the risk of mortality, productivity, and building damage level, are considered based on the outputs 
    from the simulation. In addition, probabilistic loss scenarios and outcomes are generated considering various factors, including health conditions, healthcare accessibility, and survivability, to estimate the exposure risk of individual occupants.
    \item  ``Loss valuation" is to quantify the damage during the event to express in monetary terms including VSL, medical costs, productivity decrease costs, and building maintenance and repair costs. 
\end{itemize}

This paper focuses on the last two stages to identify the damage and estimate the economic value associated with the damage based on the simulation results from exposure analysis. Since the use-case chosen for the demonstration of the framework aims at emulating the Texas freeze event during winter storm Uri, some of the functional elements are tuned towards outages during extreme cold temperatures. However, it should be noted the valuation framework is applicable to both extreme hot and cold temperature-related events.
\vspace{-0.25cm}

\subsection{Preliminaries}
In this work, probabilistic scenarios and outcomes are generated to estimate occupant loss (i.e., mortality). The mortality rate is the number of deaths in a given population over a specified period of time. 
The mortality rate during extreme temperature conditions can be affected by several factors such as age, pre-existing health conditions, socioeconomic status, duration of the event, transportation availability, and more \cite{Rocklov2014, wang2016, ONeill2003}. 
For converting the damage to monetary terms, several assumptions are made.

First, it is assumed that severe health problems related to extreme temperatures can lead to mortality. During cold temperature events, these can be severe hypothermia or exacerbation of pre-existing illnesses with respiratory and cardiovascular failure. Mild health problems, such as respiratory infections, are not considered. 

Second, during outages, occupants can take precautions to prevent temperature-related diseases by wearing appropriate clothing and using heaters/coolers that do not require electricity. 
Individuals may also be able to access medical care or community centers (\textit{e.g.}, warming stations) when needed. This work currently assumes that these do not remove the increased mortality risk during extreme temperature events.

Third, the effect of age on the mortality rate is not considered because of conflicts in previous research. An analysis of the temperature and mortality relation in 11 US cities found that the cold-related mortality association was stronger among those over the age of 65 years \cite{Curriero2002}. In contrast, another study found stronger cold associations for those under the age of 65 years, but heat effects did not vary by age \cite{ONeill2003}. 

Fourth, transportation availability is not considered in this work. Although extreme temperature-related outage events can create hazardous road conditions and disrupt public transportation systems, 
our simulation does not include models to consider transportation from individual homes to hospitals/community centers. In addition, we have insufficient data sources for modeling accessibility and capacity to fully consider transportation availability.

Fifth, the socioeconomic status of customers is associated with their susceptibility to the impacts of service disruptions as minority and low-income populations are less likely to have resources (house insulation level and insurance availability) to withstand extreme temperatures \cite{lee2022community}.  
Detailed income level was not considered in this work due to the complexity of mapping individual occupants and housing attributes to income level. 


With the above assumptions and justifications, we have carefully identified the necessary variables in determining customer interruption costs in the following subsections.

\section{Functional models for loss estimation} \label{sec:function_models}

\begin{figure}[!t]
\centering
\includegraphics[width=0.5\textwidth]{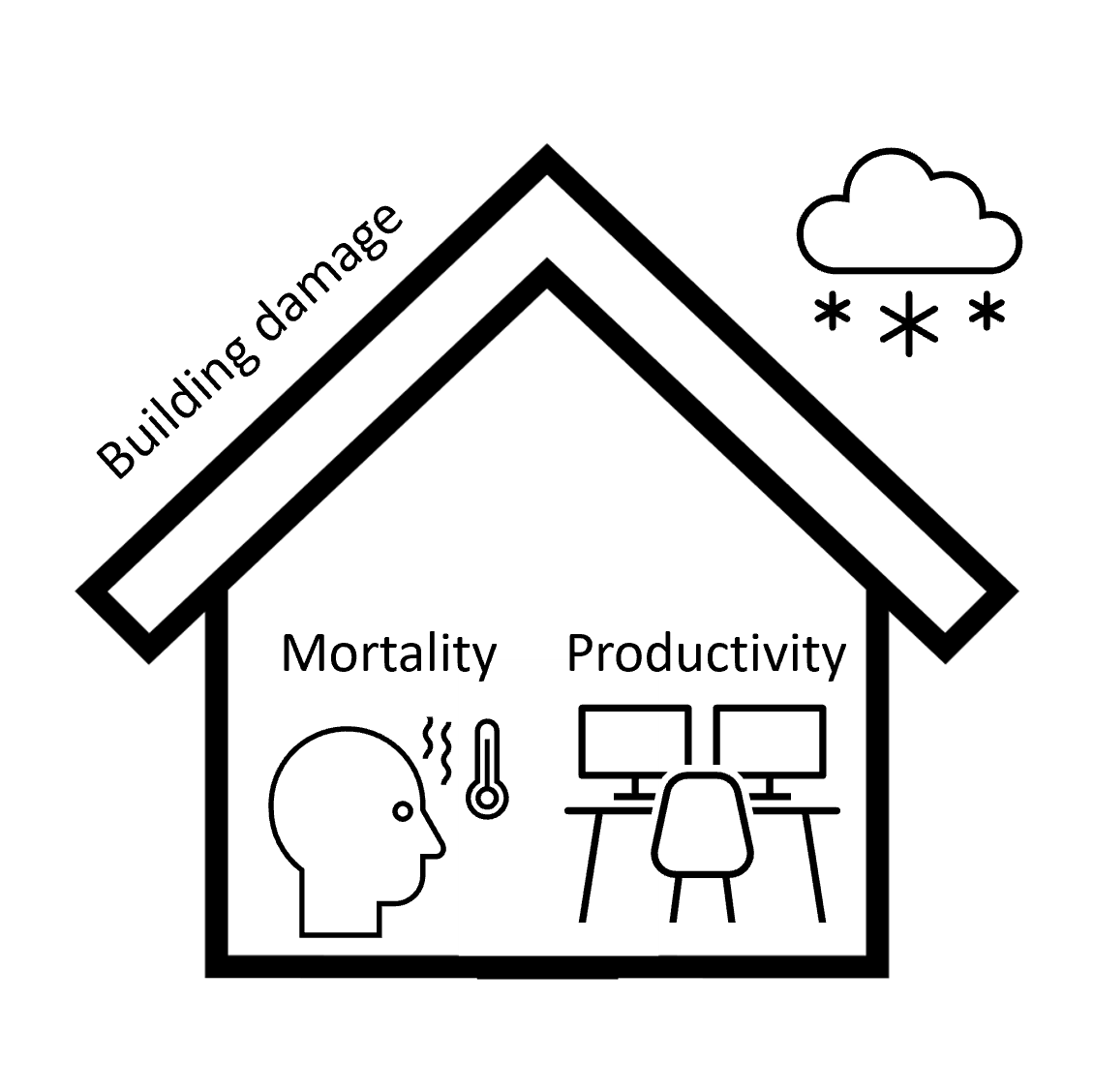}
 \vspace{-0.25 cm}
\caption{Damages associated with extreme cold weather-related outage}
\label{fig:damages}
 \vspace{-0.5 cm}
\end{figure}

Fig. \ref{fig:damages} identifies the damage and loss related to outages during extreme conditions that are considered in this work. In this subsection, the damage function models of mortality, productivity, and property damage are described with assumptions and justifications to evaluate the severity of the events, as a means of quantifying the risk of potential damage.  
 
\subsection{Mortality Model} \label{sec:MortalityModel}
\begin{figure}[!t]
\centering
\includegraphics[width=0.6\textwidth]{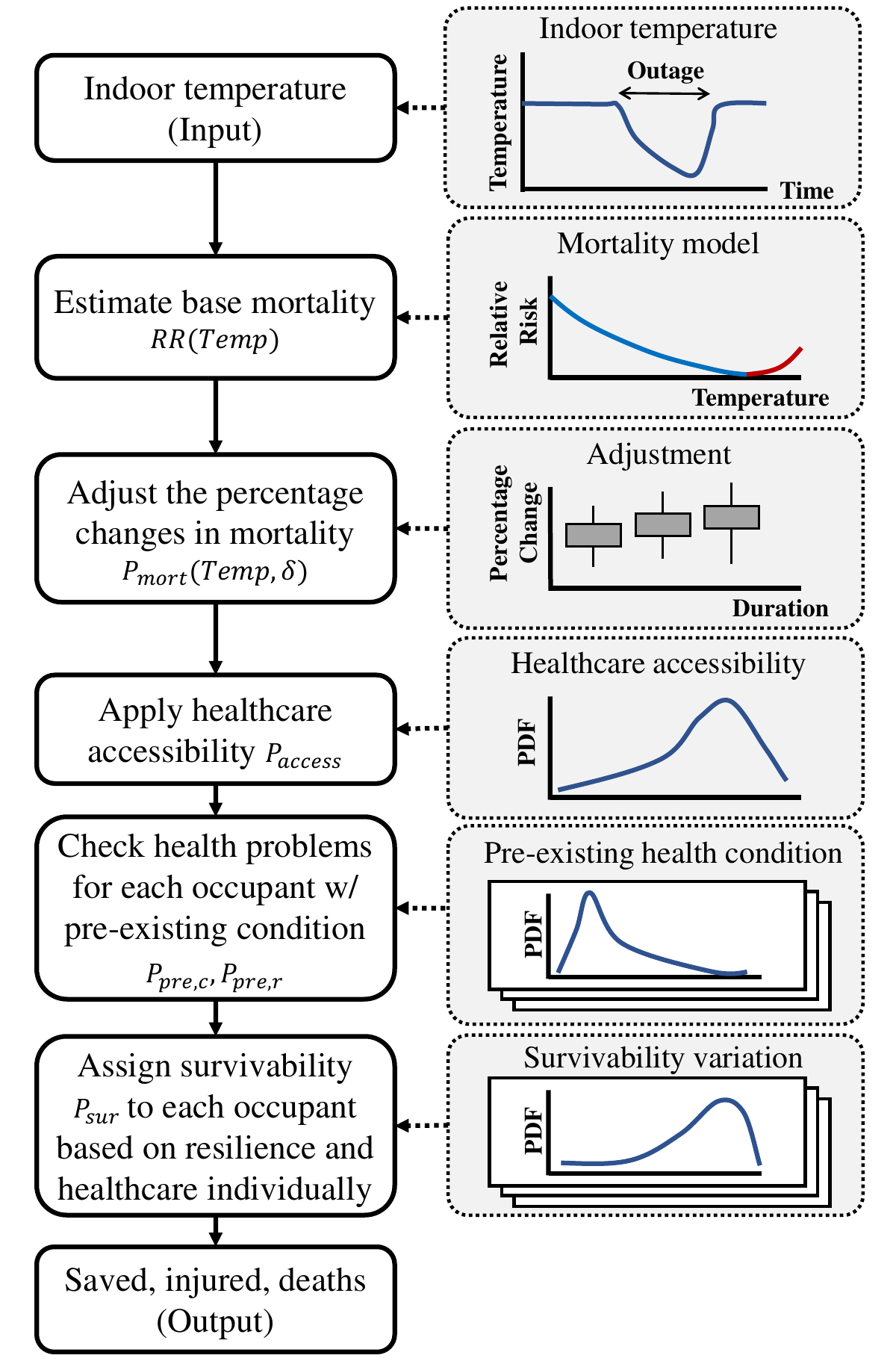}
 \vspace{-0.25 cm}
\caption{Flow of the probability-based method for the mortality estimation}
\label{fig:mortalitymodel}
  \vspace{-0.25 cm}
\end{figure}

The mortality rate is modeled using the function of relative risk (\textit{RR}) of mortality to indoor temperature during extreme temperature-related outages. As shown in Fig. \ref{fig:mortalitymodel}, the base \textit{RR} for occupants 
is estimated using the indoor operating conditions. This \textit{RR} is then used to estimate mortality based 
on the duration of exposure, healthcare accessibility, pre-existing health conditions, and survivability of individuals through probabilistic outcomes as illustrated in Fig. \ref{fig:mortalitymodel}.
In this way, the mortality estimation takes into account the building characteristics of customer premises with other medical factors.

\subsubsection{Function of Mortality to Cold Temperature}

\begin{figure}[!t]
\centering
\includegraphics[width=0.65\textwidth]{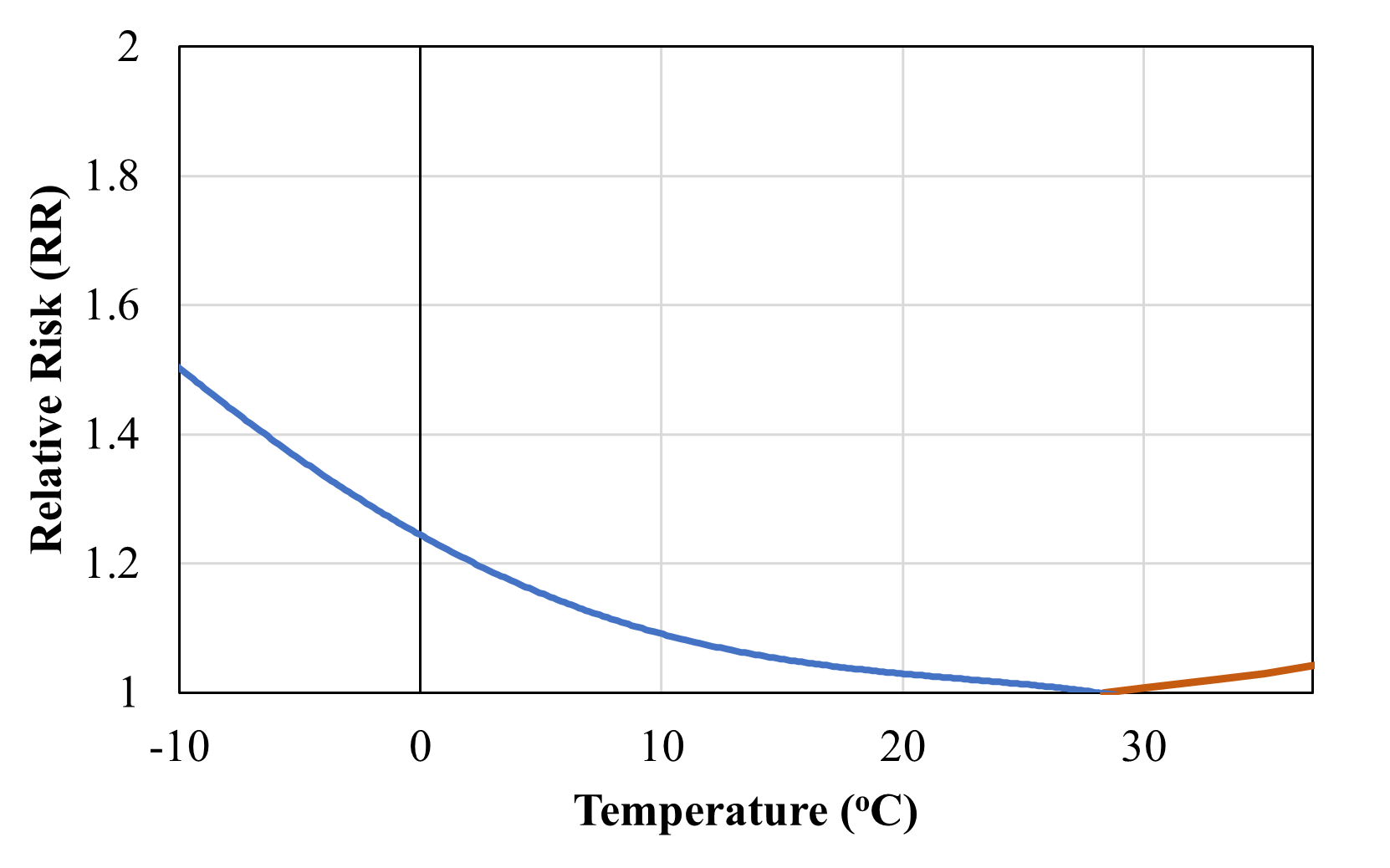}
 \vspace{-0.25 cm}
\caption{\textcolor{black}{Function of relative risk of mortality to temperature} \cite{Gasparrini2015}}
\label{fig:mortfunctemp}
 \vspace{-0.5 cm}
\end{figure}

Fig. \ref{fig:mortfunctemp} shows a function of relative risk of mortality based on the temperature. This framework utilizes the specific location-based curve, as referenced from \cite{Gasparrini2015}. Fig. \ref{fig:mortfunctemp} shows the curve for Austin, TX, and this framework is capable of assessing the impact of temperature on mortality risk in particular areas. 
It can be formulated with the indoor temperature during an extreme temperature-related outage using a regression model as shown in \eqref{eq:RRtemp}.
\begin{align}
    RR(Temp_{b,t}) = a_1 Temp_{b,t}^4 + a_2 Temp_{b,t}^3 + a_3 Temp_{b,t}^2 + a_4 Temp_{b,t} + a_5 \label{eq:RRtemp}
\end{align}
where $RR(Temp_{b,t})$ is relative risk level of mortality based on the temperature $Temp_{b,t}$ of a building $b$ at time $t$.

In addition to that, the influence of the duration of the exposure 
can be included to adjust the mortality rate as longer and more severe cold waves can lead to higher mortality rates \cite{wang2016}.
Then, the base mortality rate is estimated using \eqref{eq:mortality}. 
\begin{align}
    P_{mort,b}(Temp_{b},\delta) = (\overline{RR(Temp_{b})}-1) + \delta \label{eq:mortality}
\end{align}

where $\overline{RR(Temp_{b})}$ is the average relative risk of mortality by temperature throughout the outage event, and $\delta$ is the percentage changes in mortality considering the duration and severity of the event.

\subsubsection{Healthcare Accessibility}
Healthcare accessibility ($P_{access}$) can be determined by various factors, such as transportation availability, distance, income-level, and so on. Since this paper does not consider transportation availability and income-levels,  a truncated normal distribution (with no values below zero), based on a particular mean value adjusted with respect to the insurance rate of the region under consideration, was leveraged to identify the healthcare accessibility of individuals. Further details of the parameters used for the distribution are given in \ref{appendix}. Note that this is a rough estimate, and the actual healthcare accessibility can vary widely. This approach was selected to consider some reasonable numbers with available information and simplify the model. 

\subsubsection{Health Impacts}
Exposure to extreme weather during a power outage can lead to several adverse health impacts for people. 
The health impacts considered in this work are frostbite, hypothermia, respiratory problems, and heart conditions.
To model these health impacts, the probabilities of pre-existing health conditions were utilized to determine the individuals who are at higher risk of mortality. Previous research has shown that the most significant mortality outcomes of cold weather-related health issues are cardiovascular and respiratory-related \cite{Ha2009}. Thus, this work considered the percentages of occupants with cardiovascular disease ($P_{pre,c}$) and respiratory problems ($P_{pre,r}$). The percentage of occupants with pre-existing health conditions in each household was modeled using a truncated normal distribution with mean values based on health survey data for the chosen region (see \ref{appendix}). The truncated normal distribution was chosen for simplification and target value. Once the probabilities for people with pre-existing health conditions were determined, it was assumed that the rest of the individuals would still be vulnerable to frostbite or hypothermia during the cold weather-related outage.

\subsubsection{Survivability}
In this work, survivability is attributed to individuals at risk of mortality and is related to occupant's own resilience to recover from health impacts experienced during the extreme cold weather outage events. 
Survivability varies depending on which health impacts people are suffering and their geographic location in relation to resources for dealing with those issues. 
For example, 
patients who visited hospitals with severe frostbite and hypothermia had a survivability rate that was approximately 42.5\% higher when compared to those who were not treated, as reported in \cite{Ko2019}. 

During an outage caused by extreme weather, even hospital survivability can be affected by various factors, including hospital capacity. 
This would differ for every hospital and healthcare facility depending on the location, size, and resources. However, it is challenging to estimate the likelihood. 

Survivability ($P_{sur}$) is modeled considering their health impacts and healthcare accessibility using the truncated normal distribution with statistical parameters (e.g., mean, standard deviation, etc.) based on earlier research and available resources (see \ref{appendix}). Truncated normal distribution in our model allows for reasonable estimates using the available data while keeping the model simple.
With the probability models of health impacts and healthcare accessibility for individuals, we can map out possible health issues that people may suffer during extreme cold weather events and where they suffer/survive from the health issues. People can either stay at home or visit appropriate healthcare to recover from the health issues caused by the cold weather during an outage.


\subsection{Productivity Model}
\begin{figure}[!t]
\centering
\includegraphics[width=0.65\textwidth]{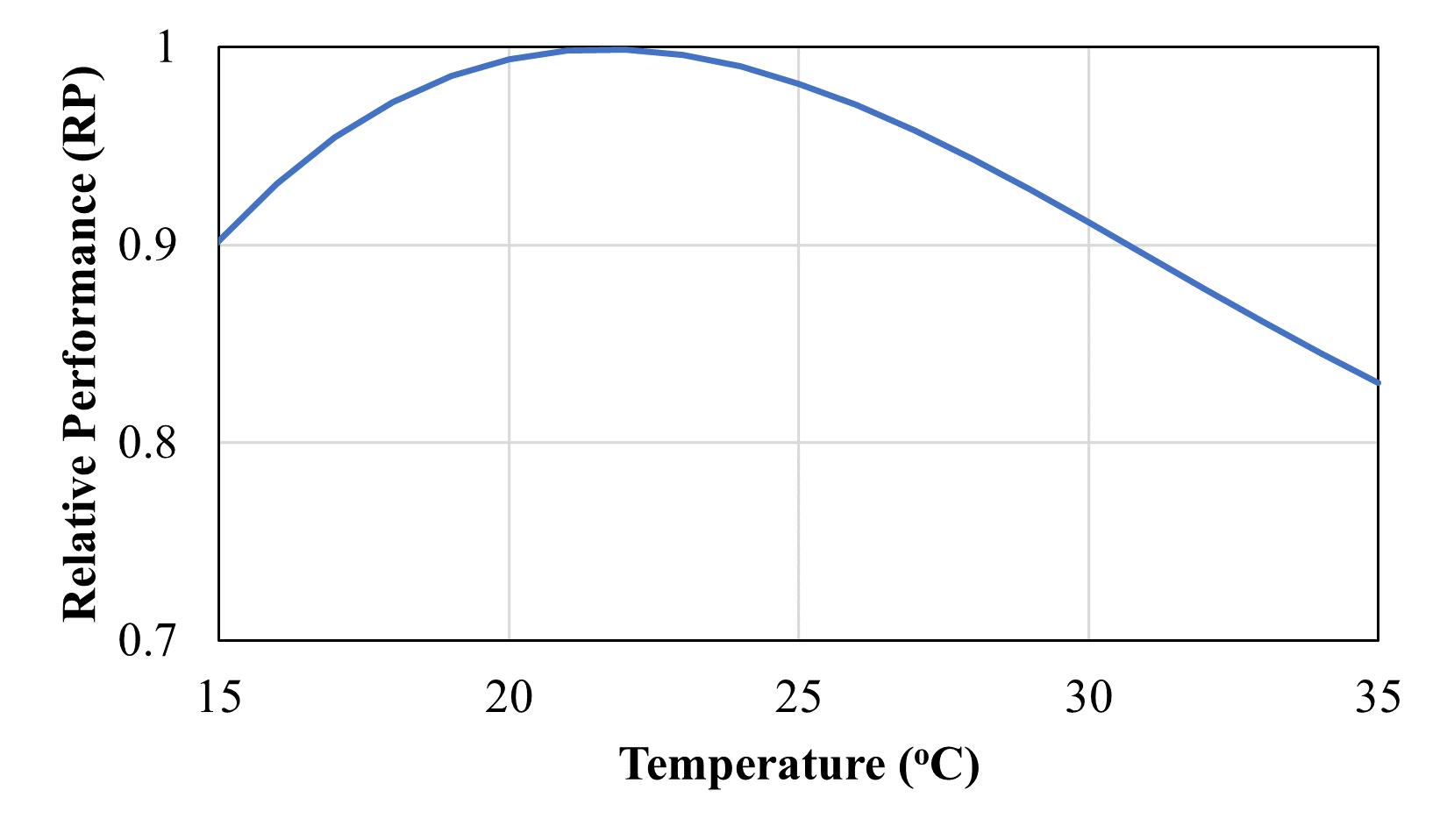}
 \vspace{-0.25 cm}
\caption{Relative performance level by indoor temperature \cite{Seppanen2006}}
\label{fig:productivity}
 \vspace{-0.25 cm}
\end{figure}

Productivity can be reduced due to discomfort caused by room temperature as in Fig. \ref{fig:productivity}. A previous study found that as the temperature deviates from the optimal temperature range (which is between 21°C to 23.5°C or 70°F to 74°F), there is a corresponding decrease in productivity levels \cite{Seppanen2006}. 
Productivity decrease due to room temperature deviation can be estimated with the following equation \cite{Seppanen2006}. 
\begin{align}
     P_{prod,b,t}(Temp_{b,t}) = d_1 Temp_{b,t}^3 + d_2 Temp_{b,t}^2 + d_3 Temp_{b,t} +d_4 \label{eq:prod}
 \end{align}
where $P_{prod,b,t}$ is the relative performance of productivity by the indoor temperature of a building $b$ at time $t$.
To analyze the impact of extreme temperatures on productivity under various job categories and power availability conditions, during the outage period, the productivity level is assessed only for non-power requirement jobs. For power requirement jobs, the assessment is conducted specifically when intermittent power is available. 

\subsection{Property Damage Model}
Extreme weather can cause damage to buildings \cite{Koci2017}. 
For instance, during extreme cold events, when temperatures drop below freezing, pipes can freeze and burst, leading to water damage inside the building. An index derived from temperature and moisture content is used to identify if cold temperatures damage the building \cite{Koci2017}.
The following Winter Index (WI) is calculated when both moisture and temperature fall outside the critical levels.
\begin{equation}
\small
     WI_t = (T_L - T_t) (RH_t - RH_L) [(T_t < T_L, RH_t > RH_L] \label{eq:WI}
 \end{equation}
where $T_L$ and $RH_L$ are critical values of temperature and moisture content and $T_t$ and $RH_t$ are temperature and moisture content at time $t$, respectively. WI is calculated only when both $T_t < T_L$ and $RH_t > RH_L$ are satisfied.

Since the WI in this work is calculated based on the indoor air temperature of each building, we can differentiate the damage among the buildings in the population. By utilizing the above index, we can assign a random binary number (e.g., 0 or 1) to represent the probability $P_{damage,b}$ of requiring repairs for a specific building $b$.


\section{Loss Valuation}\label{sec:loss_valuation}

\subsection{Value of Statistic Life (VSL)}
VSL is a commonly used measure that assigns a monetary value to an increased risk of mortality. It is important to note that the VSL is not a measure of the intrinsic value of a specific human life. Rather, it is a measure of how much people are willing to pay to reduce the risk of mortality.
Several studies on VSL provide different estimates for different populations and contexts. For example, a study provides a review of VSL estimates from different countries, which range from about \$0.1-\$15 million \cite{Viscusi2003}. For COVID-19-related deaths, the VSL is estimated to be \$11.5 million \cite{Hammitt2020}. Federal Emergency Management Agency (FEMA) also provides guidance on conducting benefit-cost analysis that includes estimating the VSL of \$11.6 million \cite{FEMA}. The U.S. Department of Transportation provides specific values of \$11.8 million for the VSL based on the methodology adopted in their guidance \cite{DOT}.

In this work, the total number of deaths is estimated using the probabilistic loss model in Fig. \ref{fig:mortalitymodel} using \eqref{eq:Ndeath}. 
The total VSL $C_{vsl}$ during the event is calculated based on the $VSL$ per life as shown in \eqref{eq:Cvsl}.
\begin{align}
     &N_{death} = f(P_{mort}(Temp,\delta), P_{access}, P_{pre}, P_{sur}) \label{eq:Ndeath}\\
     &C_{vsl} = \sum_{b=0}^{B} N_{death,b} VSL \label{eq:Cvsl}
 \end{align}
where $N_{death}$ is the number of deaths for building $b$ estimated with the probability-based mortality model. 
 
\subsection{Medical Cost}
This work assumes the medical costs associated with health impacts are only based on the danger of mortality due to extreme weather exposure.
The number of individuals injured and recovered from health care can be estimated using the probabilistic loss model as given in \eqref{eq:Ninjured} (see in Fig. \ref{fig:mortalitymodel}). 

The medical cost can vary depending on the specific health problems that people experience as well as the length of hospital stay \cite{Boucher2022}.
To account for severity, the base mortality rate $P_{mort}(Temp,\delta)$ is used in the linear cost model with a cost range specific to health issues. 
Additionally, the probability of health insurance coverage is considered to estimate the total medical costs $C_{medical}$ as in \eqref{eq:Cmedical}.
\begin{align}
     &N_{injured} = f(P_{mort}(Temp,\delta), P_{access}, P_{pre}, P_{sur}) \label{eq:Ninjured}\\
     &C_{medical} = f(N_{injured}, P_{mort}(Temp,\delta), P_{heal,ins}) \label{eq:Cmedical}
 \end{align} 
 
 \subsection{Productivity Decrease Costs}
 Productivity decrease cost is calculated by multiplying the reduction in productivity by the hourly wage rate of the affected workers. The resulting cost represents the economic impact of potentially reduced productivity due to discomfort caused by room temperature. As mentioned earlier, the productivity-related cost is estimated for non-power requirement jobs during the outage period, and for power requirement jobs, it is conducted specifically when intermittent power is available. Residential sector productivity is historically not monetized, with the recent increase in working from home, this is a cost that a customer may incur during an outage and be appropriate to include in the valuation analysis. Some NEI estimates exist for the value of comfort in residential buildings. However, existing estimates for comfort non-energy impacts are typically associated with energy savings from home weatherization programs and are not easily generalized to this research.  
The total productivity decrease costs $C_{prod}$ can be calculated during the affected hours as follows. 
 \begin{align}
      C_{prod} = \sum_{b} \sum_t N_{work,b}  (1- P_{prod,b,t}(Temp_{b,t})) W_{avg} \label{eq:Cprod}
 \end{align}
where $N_{work,b}$ is the number of workers in building $b$ and $W_{avg}$ is the average hourly wage rate of the affected workers.

\subsection{Building Repair Cost}

The total repair costs $C_{build}$ can be calculated based on the building insurance $P_{build,ins}$, labor costs $C_{labor}$, and any material or equipment costs $C_{item}$ as in \eqref{eq:Cbuild}. To account for severity, the sum of $WI$ over the outage period is used in the linear cost model with a minimum and maximum cost range specific to the repair item as in \eqref{eq:sumWI_cost_map}. 
\begin{align}
     & C_{build} = f(P_{damage} P_{build,ins}, C_{labor}, C_{item}) \label{eq:Cbuild} \\
      &C_{item} = \alpha_{\text{min}} + (\alpha_{\text{max}} - \alpha_{\text{min}}) \left(\frac{\sum_t WI_t}{\beta_{\text{WI}}}\right) \label{eq:sumWI_cost_map}
 \end{align}
 where $\alpha$ is the cost specific to a particular repair item and $\beta_{\text{WI}}$ is the maximum accumulated value of WI during the event. 
 
\subsection{Power Interruption Cost}
This framework also includes the general power interruption costs for the customers based on the customer types \cite{ICEmodel} as in \eqref{large_cic} - \eqref{res_cic}. To enhance the limitation of \cite{ICEmodel}, a scalability factor is also included to estimate the power interruption costs for outages beyond 16 hours. 
\begin{align}
& \text{For large and medium C\&I}, \nonumber \\ 
& C_{cic} = f(e_{avg}, \Delta t_o, szn, ind) \label{large_cic}\\
& \text{For small C\&I}, \nonumber \\ 
& C_{cic} = f(e_{avg},  \Delta t_o, szn, ind, backup) \label{small_cic}\\
& \text{For residential}, \nonumber  \\ 
& C_{cic} = f(e_{avg},  \Delta t_o, inc, szn) \label{res_cic}
\end{align}
where $e_{avg}$ is the average annual consumption of the customer, $\Delta t_o$ is the duration of the outage events which is determined by the actual power-off time for each building. $szn$ is the season of interruption, $ind$ denotes the customer business type, $backup$ denotes the presence of backup equipment at the building, $inc$ is the annual household income.



\subsection{Monte-Carlo simulation}
\begin{figure}
    \centering
    \includegraphics[width=0.6\linewidth]{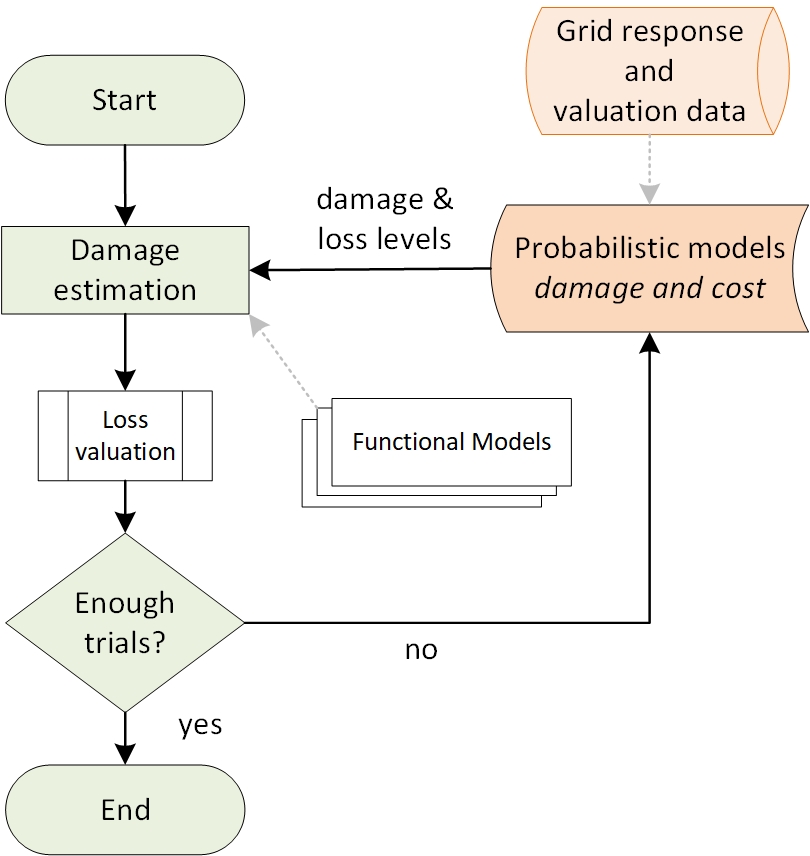}
    \caption{Monte-Carlo simulation for loss estimation and valuation.}
    \label{fig:mcs}
     \vspace{-0.25 cm}
\end{figure}

The framework is based on Monte Carlo simulation to model the probabilistic damage and cost scenarios and it requires simulation outcomes and valuation data. 
Monte Carlo simulation techniques consider the stochastic nature of the impact of extreme temperature-related outages. In this context, it is used to generate a damage scenario using grid response and exposure analysis data (see Fig. \ref{fig:mcs}). The damage and loss levels of occupants and buildings are determined by mapping the probability distribution function. The generated damage scenario is then used to determine the total interruption cost. Several trials of Monte Carlo simulation are conducted to obtain statistically representative results. 

\section{Modeling Distribution Grid Response }\label{sec:grid_modeling}
The framework, proposed in Section \ref{sec:value_models}, presents a set of functional models and metrics for estimating the net value of customer loss due to outages during extreme temperature-related events. Towards demonstrating how the valuation framework could be implemented for damage assessments and along with quantifying the potential value of resilience enhancement due to grid modernization, a simulation-based use-case is designed to emulate the impacts of customer outages in Texas during the winter storm Uri in Feb. 2021 \cite{TexasNews}. 

The use-case design adopts an at-scale distribution grid response modeling approach to demonstrate large-scale system response to extreme temperatures \cite{hanif2022analyzing}. The response models incorporate a detailed representation of the system’s physical components (like physics-based end-user load models) along with operational parameters. Towards emulating the grid response of Texas, these models are configured based on the key characteristics of the region in consideration, which include:
\begin{itemize}[leftmargin=0.35 cm]
    \item Type of utilities (urban, suburban, and rural) - Utilized to model the backbone infrastructure (e.g., topology, equipment rating, power conversion, and delivery elements) through appropriate prototypical feeders \cite{Schneider2008taxonomy}.
     \item Geo-spatial location of the region - Utilized to model the region’s climate zone along with appropriate weather profile to emulate extreme temperature-related events. 
     \item Residential, commercial, and industrial (RCI) mix -  Utilized to model the residential and commercial customers through physics-based load models. Each residential customer is modeled through a combination of temperature-sensitive loads, based on thermostat settings, heating fuel source, floor areas, water-heater type, and other plug loads based on consumption profiles and distributions obtained from the energy consumption surveys conducted by the U.S. Energy Information Administration (EIA), further details of which are presented in \cite{DSOT}.
\end{itemize}

\begin{figure}[!t]
\centering
\includegraphics[width=\linewidth]{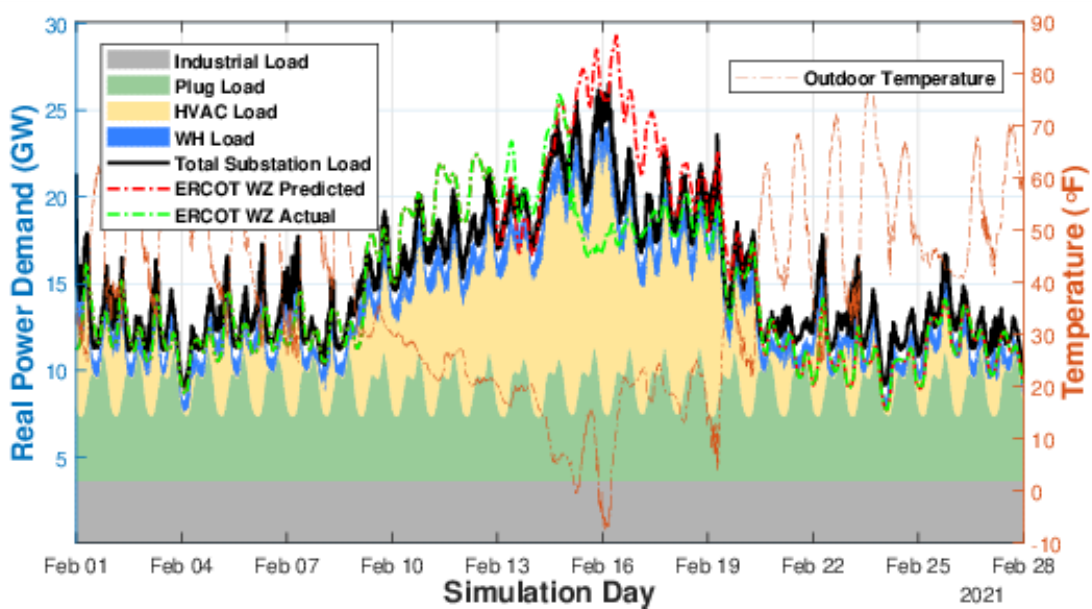}
 \vspace{-0.5 cm}
\caption{Load decomposition obtained from the grid response models for Feb. 2021 calibrated against ERCOT load data for the relevant weather zone.}
\label{fig:base_month}
 \vspace{-0.25 cm}
\end{figure}

\begin{table}[!t]
\centering
\caption{Summary of Simulated Response Models}
\label{UseCase}
\begin{tabular}{l|c|c|c}
\toprule[0.5 mm]
\hline
\multirow{1}{*}{Characteristics} & Simulated & Scaled & Actual WZ\\
\hline
Residential Customers  & {1308} & {3,075,342}& {3,075,342} \\
Commercial Customers   & {95} & {223,362}& {232,720}  \\
\hline
Total Customers   & {1403}& {3,298,704}& {3,447,423}  \\
\toprule[0.5 mm]
\hline
Average System Load &{4.95 MW}& {15.29 GW}& {15.48 GW} \\
Expected Max. System Load &{9.34 MW}& {25.61 GW}& {29.30 GW} \\
Minimum System Load &{2.30 MW}& {9.06 GW}& {7.75 GW} \\
\hline
\toprule[0.5 mm]
\end{tabular}
\end{table}

\subsection{\textBlack{Response Calibration}}
The use-case is configured to emulate the distribution grid response for the North Central (NC) weather zone (WZ) of Texas during the storm. The NC zone is one of the eight weather zones of ERCOT \cite{WZ_ercot} and experienced a heavy increase in demand during the winter storm Uri in Feb. 2021.  The NC-WZ of Texas is mostly urban in its demographics and was modeled with a distribution substation connected with two urban prototypical feeders (R5-12.47-1 and R5-12.47-2) \cite{Schneider2008taxonomy} and a scaling factor based on the business-as-usual (BAU) conditions of NC WZ of Texas. These feeders are configured with the customer mix and end-user loads for NC-Texas (based on EIA data).  Further details on the scaling factor, weather data, and calibration are presented in \cite{hanif2022analyzing}. Figure \ref{fig:base_month} presents the simulated grid response and compares it with the actual (with outages) and predicted load data (without outages) for the NC-WZ of Texas region during Feb. 2021. The actual (ERCOT WZ Actual) demand served by ERCOT during the extreme winter temperatures is taken from~\cite{ercot_hist}. The predicted value represents the estimated demand during winter storm Uri if there had been no outages and is obtained from \cite{gruber2022profitability}. Table \ref{UseCase} summarizes the simulated response models. It can be seen from Fig. \ref{fig:base_month} that the simulated response matches well with the predicted load during outage days and the actual loads during the non-outage days, respectively, presenting a baseline for implementing the valuation framework for damage assessment. 

\begin{figure}[!t]
\centering
\includegraphics[width=\linewidth]{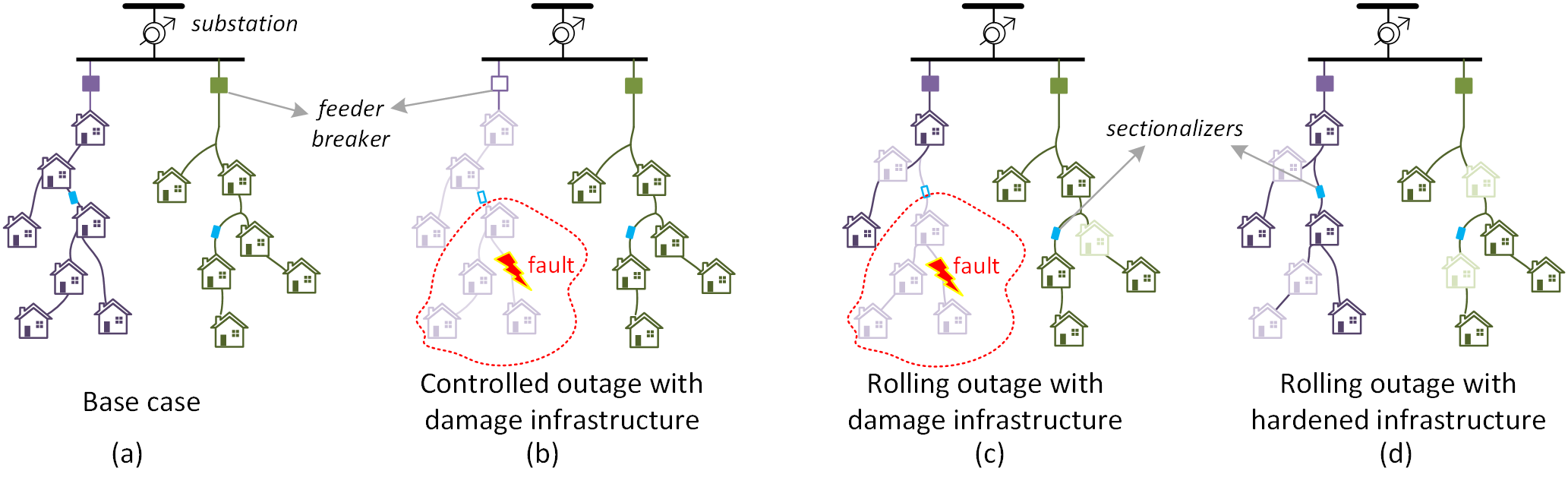}
 \vspace{-0.75 cm}
\caption{\textBlack{Overview of the outage scenarios, including damaged and hardened infrastructure. The red dotted circle represents customers impacted by infrastructure damage (i.e., sectionalizer open due to a line fault).}}
\label{fig:outage_scenarios}
 \vspace{-0.5 cm}
\end{figure}

\subsection{Infrastructure Vulnerability}\textBlack{
Extreme events such as snow and ice storms can damage power distribution infrastructure. Quantification of such damage and resilience assessment requires a detailed analysis of different components in a probabilistic framework such as i) event characterization, ii) component fragility model, iii) system performance model, and iv) repair or restoration model \cite{hou2023resilience}. Note that the primary focus of this paper is to assess the value of customer resilience by considering both the direct economic losses and non-energy impacts stemming from extreme temperature-related outages. While our emphasis lies on estimating customer resilience rather than modeling the cause of the outage (based on event assessment and component fragility as proposed in \cite{hou2023resilience}), we introduce a simulated fault condition during the extreme temperature period to account for the impact of infrastructure damage on resilience. Furthermore, it is assumed that any infrastructure damages incurred will be repaired by the end of the scarcity period, allowing us to conduct the performance comparison effectively. Fig. \ref{fig:outage_scenarios} shows a schematic of different scenarios, demonstrating how the infrastructure damage is modeled along with its impact when operated with different resilience enhancement scenarios.  
It is shown that a fault on an overhead line causes an upstream sectionalizer to open, and a few customers will be in outage condition (see dotted red circle in Fig. \ref{fig:outage_scenarios}b and \ref{fig:outage_scenarios}c). For example, when implementing a resilience strategy (\textit{e.g.}, rolling outage) with damaged infrastructure, certain customers will be in the isolated section, and the rest will be supplied in a rolling fashion based on the given availability. In this work, infrastructure damage is simulated using line faults that caused $3.4\%$ of customers to be impacted within the feeder. Four different scenarios depicted in Fig. \ref{fig:outage_scenarios} are simulated using the distribution grid models, and the grid response for each scenario will be discussed in the demonstration section.}

\section{Demonstration}
In this section, we demonstrate the proposed valuation framework using the simulation-based use case described earlier in Section \ref{sec:grid_modeling}. First, we present three scenarios specifically designed to assess the potential economic impact of the extreme cold temperature-related outage. Subsequently, we showcase simulation findings that emphasize the exposure analysis and valuation for each of these three scenarios. Detailed parameters for the statistical distributions used to develop the valuation as defined in Section \ref{sec:function_models} and Section \ref{sec:loss_valuation} are described in \ref{appendix}.

\subsection{Scenarios}
    \subsubsection{Base Case}  This case refers to a normal operating condition where the electricity demands of all connected loads are adequately met. The grid response for this case is shown in Fig. \ref{fig:base_month}. In the base case, the heating, ventilation, and air-conditioning (HVAC) systems operate normally, and it is assumed that there is no outage caused by extremely cold weather. 

    \begin{figure}[!t]
    \centering
    \includegraphics[width=\linewidth]{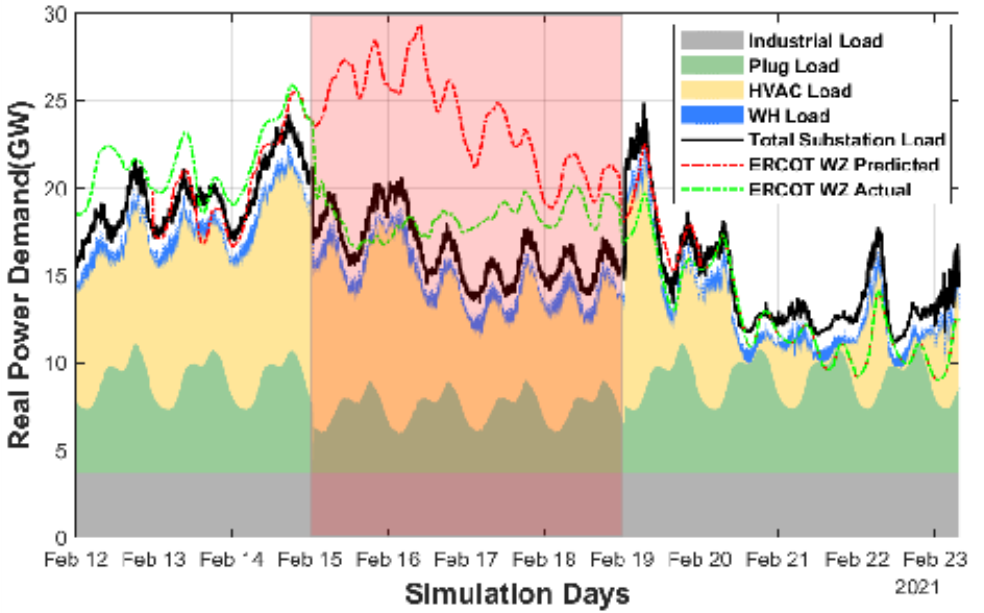}
     \vspace{-0.75 cm}
    \caption{\textBlack{Simulated Grid response for the weather zone on implementing \textit{Controlled Outages (CO)} during extreme temperature events in Feb. 2021.}}
    \label{fig:controlled_outage}
     \vspace{-0.75 cm}
    \end{figure}
    
    \subsubsection{Controlled Outage with Damaged Infrastructure (CO)}  In this case, we simulate a scarcity scenario where \textBlack{even after isolating the faulted section of the system}, the operator is unable to meet the electricity demand of the remaining sections of the feeder due to various reasons, such as high demand, generation shortages, or equipment failures. The controlled outage case emulates a strategy by utility companies to selectively switch off distribution feeders during emergencies as one of the simplest ways to reduce load. While there is a growing practice of managed power outages during severe weather events, limited evidence is available to assess how these controlled outages and subsequent power restoration efforts are executed fairly and equitably \cite{bolin2018race}. 
    To illustrate the spatio-temporal patterns of impacts on specific customer groups, we orchestrated a simulated controlled outage \textBlack{by switching off one of the distribution circuits (as shown in Fig. \ref{fig:outage_scenarios}b)} to emulate the customers in Texas who lost power during the winter storm Uri. Fig. \ref{fig:controlled_outage} illustrates how the simulated grid's response closely adhered to the ERCOT WZ's actual load profile  
    In this scenario, the customers experiencing power outages cannot operate HVAC systems to maintain thermal comfort inside the buildings due to the outage. 
    

    \textBlack{\subsubsection{Rolling Outage with Damaged Infrastructure  (RO-DI)}}   
    \begin{figure}[!t]
    \centering
    \includegraphics[width=\linewidth]{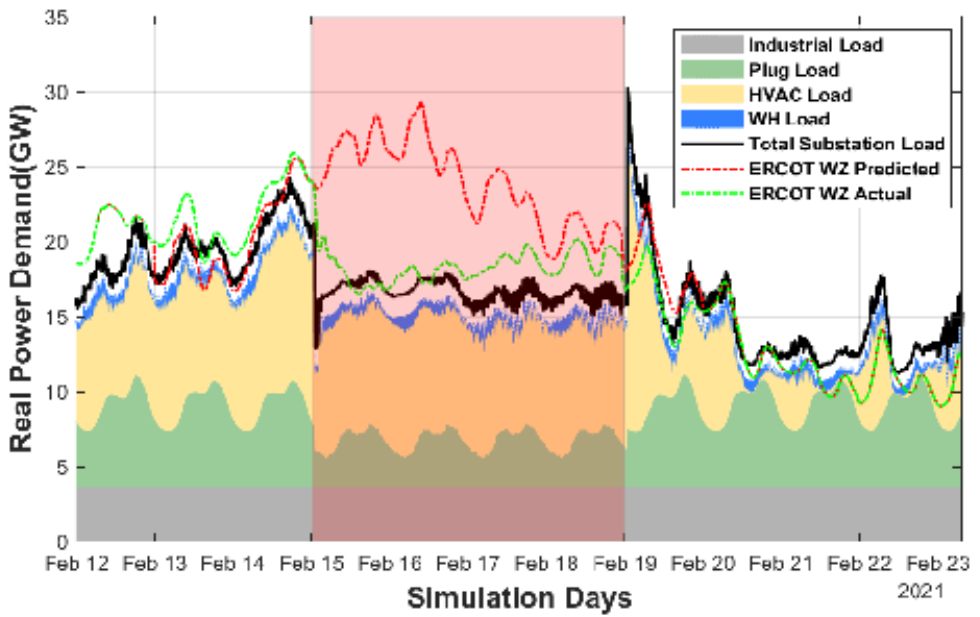}
    \vspace{-0.75 cm}
    \caption{\textBlack{Simulated Grid response for the weather zone on implementing \textit{Rolling Outages with Damaged Infrastructure (RO-DI)} during extreme temperature events in Feb. 2021.}}
    \label{fig:rolling_outage_with_infra}
    \end{figure}
    \textBlack{The rolling outage approach to deal with scarcity entails the DSO  examining historical customer data to gain insights into consumption patterns. With this knowledge, the DSO can effectively determine rolling outage scenarios, identifying a specific group of customers for remote disconnection based on availability during scarcity situations. This scenario implements rolling outages as a resilience enhancement strategy considering the infrastructure damaged during the extreme event (as shown in Fig. \ref{fig:outage_scenarios}c). Therefore, in this scenario, the customers isolated within the faulted section of the system won't be serviced throughout the event duration. Since the commercial and industrial customers have high interruption costs, in this case, the residential customers were segregated into three groups based on their past consumption, and they are sequentially supplied such that overall demand aligns with or remains within the actual demand (with outages) of the ERCOT WZ's  (see Fig. \ref{fig:rolling_outage_with_infra}).} For demonstration purposes, the rolling outages are scheduled hourly. Consequently, an individual customer experiences a service interruption lasting a maximum of two hours. A coordinated RO ensures that customers only face service interruptions for a portion of the total outage period, effectively reducing the maximum thermal stress experienced by them \cite{10262165}. Note that the group of customers to disconnect at a given time depends on the curtailment requirement and designing sophisticated RO schemes is not within the scope of this paper. 
    
\begin{figure}[!t]
    \centering
    \includegraphics[width=\linewidth]{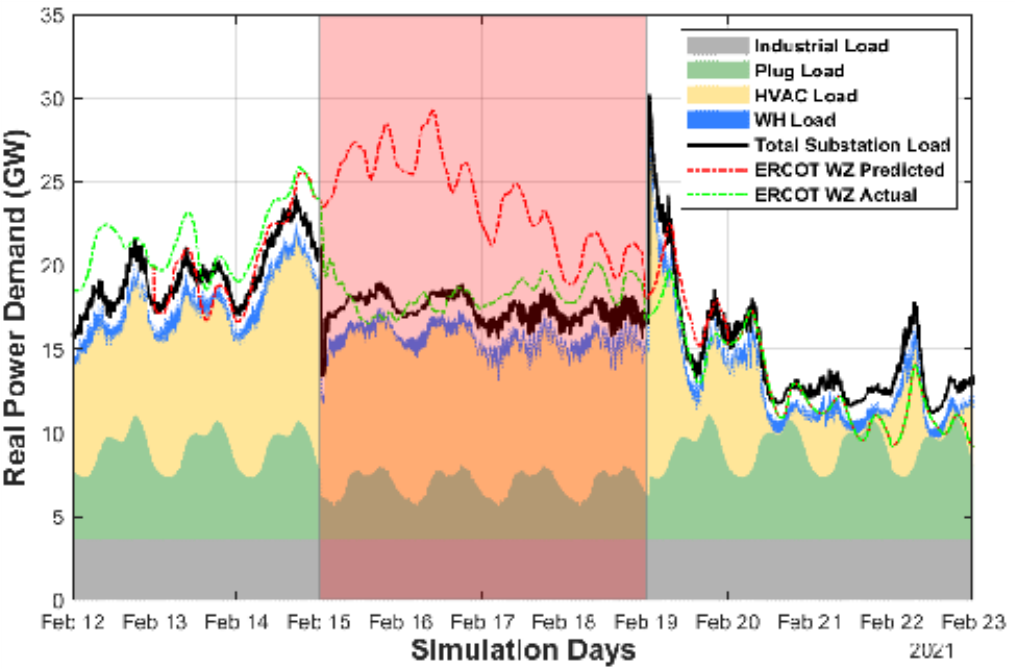}
     \vspace{-0.75 cm}
    \caption{\textBlack{Simulated Grid response for the weather zone on implementing \textit{Rolling Outages with Hardened Infrastructure (RO-HI)} during extreme temperature events in Feb. 2021.}}
    \label{fig:outage_rolling}
     \vspace{-0.75 cm}
    \end{figure}

    \textBlack{\subsubsection{Rolling Outage with Hardened Infrastructure (RO-HI)} To comparatively evaluate the impact of resilience strategies along with infrastructure upgrades, this scenario implements rolling outages as a resilience enhancement strategy considering hardened infrastructure. Therefore, this scenario doesn't consider isolation due to faulted lines, thereby enabling all customers to be sequentially serviced throughout the event duration (as shown in Fig. \ref{fig:outage_scenarios}d). Similar to the RO-DI case, the rolling outage strategy segregates the residential customers into three groups based on their past consumption and they are sequentially supplied (on an hourly basis) such that overall demand aligns with or remains within the actual demand (with outages) of the ERCOT WZ's  (see Fig. \ref{fig:outage_rolling}).}

\subsection{Results}

\subsubsection{Exposure Analysis}
\begin{figure}[ht]
\centering
\includegraphics[width=0.75\textwidth]{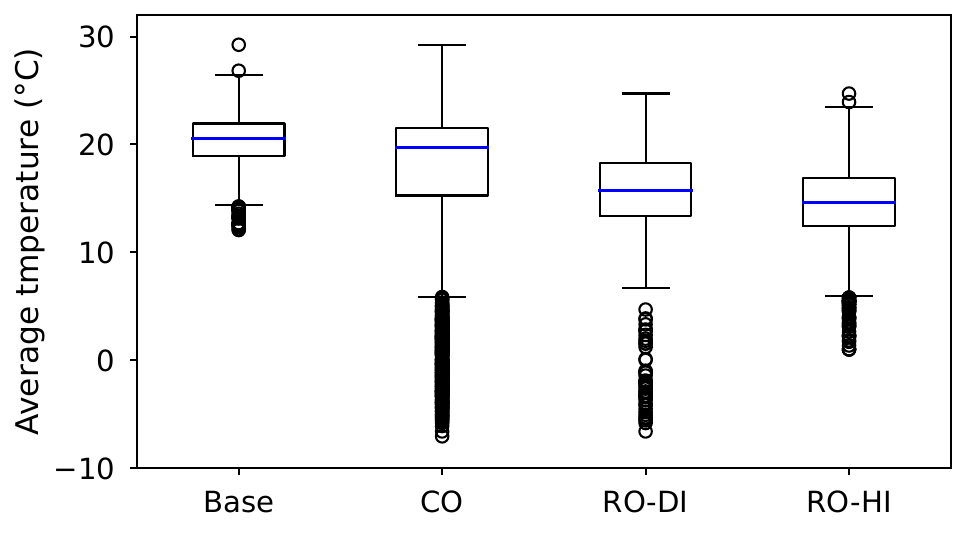}
 \vspace{-0.25 cm}
\caption{\textcolor{black}{\textBlack{Average indoor temperature of customers during the outage period (Feb $15^{th}$ - Feb $19^{th}$)}}}
\label{fig:avg_temperature_boxplot}
\end{figure}

During the outage period, the average indoor temperatures of the population across different cases are shown in Fig. \ref{fig:avg_temperature_boxplot}. \textBlack{The median values are highlighted in blue. It is shown that the median temperature of the `CO' case is close to that of the `Base' case at around $19.7^\circ$C, however, there are lots of outliers even below $-5^\circ$C in the `CO' case.} This explains that with this scheme, outaged customers could experience a significant drop in indoor temperature due to the power outage, while some customers maintain their indoor thermal comfort during the time. 
\textBlack{The `RO-DI' and `RO-HI' cases had lower median indoor temperatures (i.e., $15.7^\circ$C and $14.6^\circ$C) compared to the `CO' case. However, the `RO-HI' case did not have extreme outliers, unlike the `CO' case. Conversely, the `RO-DI' case had extreme outliers similar to the `CO' case due to damaged infrastructure, which isolated certain customers and prevented them from fully benefiting from a more intelligent outage management scheme.
This result suggests that intermittent power during the outage can help maintain the indoor temperature at a more stable level. Furthermore, it confirms that not just the outage management schemes but also the infrastructure damages have a closely related impact on the indoor temperatures of customers during the outage period. }

\begin{figure}[!t]
\centering
\includegraphics[width=0.8\textwidth]{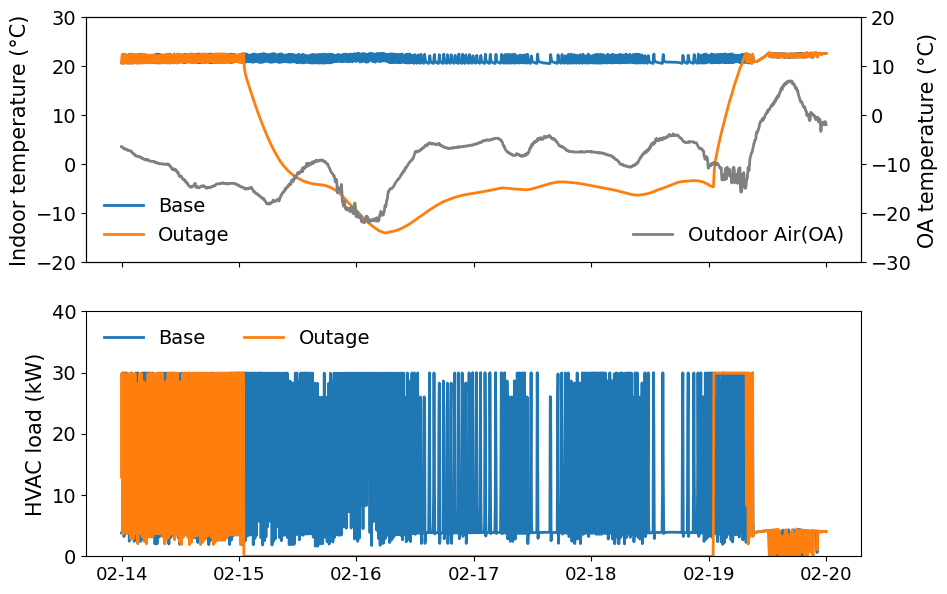}
 \vspace{-0.25 cm}
\caption{\textcolor{black}{Indoor temperature comparison between \textBlack{‘Base’ and ‘CO’ cases} in poorly insulated house during the outage (Feb $15^{th}$ - Feb $19^{th}$)}}
\label{fig:poorinsul_temp}
 \vspace{-0.25 cm}
\end{figure}

\begin{figure}[!t]
\centering
\includegraphics[width=0.8\textwidth]{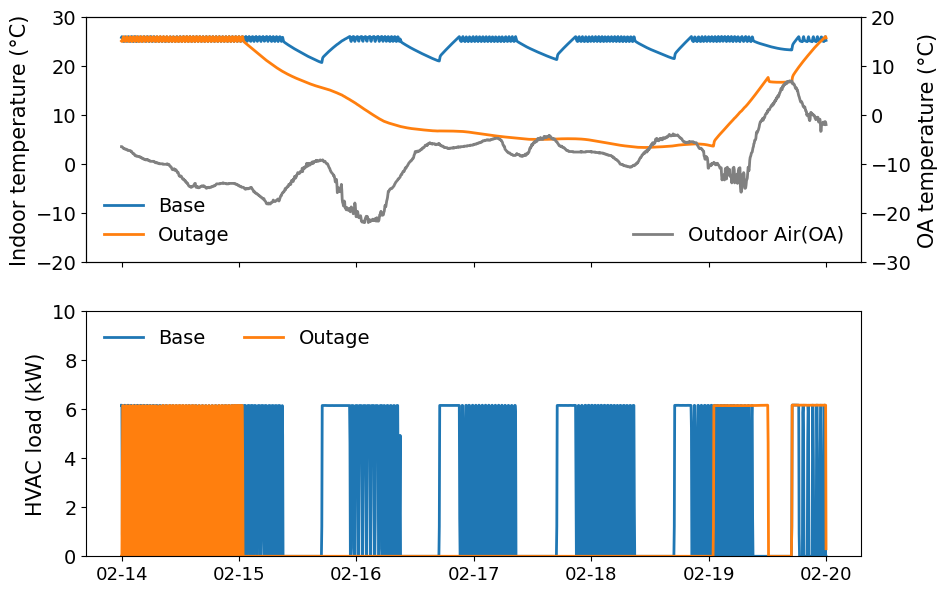}
\vspace{-0.25 cm}
\caption{\textcolor{black}{Indoor temperature comparison between ‘Base’ and ‘CO’ in well-insulated house during the outage  (Feb $15^{th}$ - Feb $19^{th}$)}}
\label{fig:goodinsul_temp}
 \vspace{-0.5 cm}
\end{figure}

Even within the same case in Fig. \ref{fig:avg_temperature_boxplot}, the indoor temperature varies depending on the building characteristics. To delve deeper into the building's performance based on its insulation properties,  Figures \ref{fig:poorinsul_temp} and \ref{fig:goodinsul_temp} show a comparative analysis of indoor air temperatures between the `Base' and \textBlack{`CO'} cases. Each figure includes two subplots: the top subplot illustrates the indoor air temperature comparison, while the bottom subplot shows the HVAC load over the time period.
In a poorly insulated house, as in Figure \ref{fig:poorinsul_temp}, the indoor air temperature can drop to \textcolor{black}{$-14^\circ$C} due to extensive heat losses through the building envelope with inadequate insulation during the outage. In contrast, Figure \ref{fig:goodinsul_temp} showcases a well-insulated house, where the indoor air temperature drops relatively slower and gently compared to the poorly insulated house. Moreover, the indoor temperature remained above \textcolor{black}{$4.4^\circ$C}, even though the HVAC system was completely off for 4 days with extreme cold weather outside.

\begin{figure}[!h]
\centering
\makebox[\textwidth][c]{\includegraphics[width=0.7\textwidth]{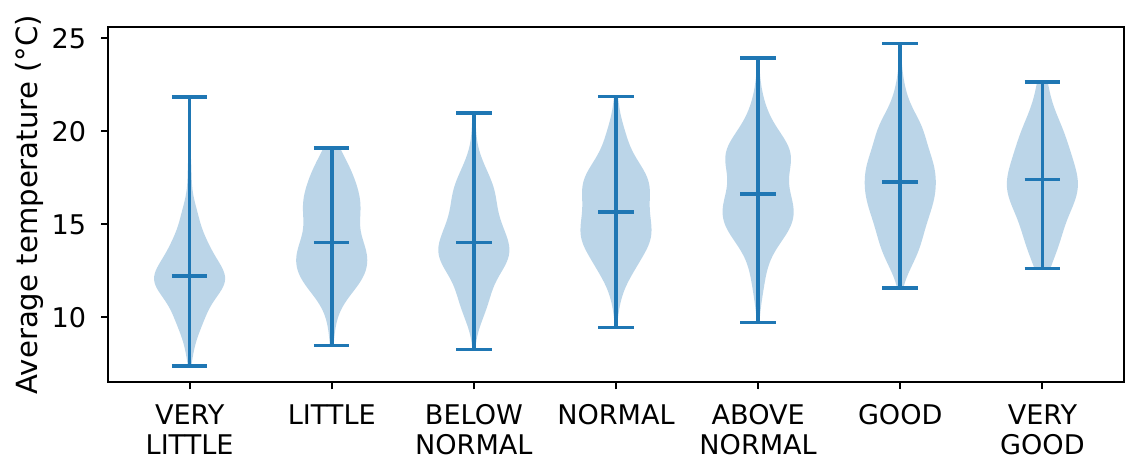}}
 \vspace{-0.5 cm}
\caption{\textcolor{black}{Average temperature by building insulation level of residential buildings in `RO-HI' case}}
\label{fig:building_thermal_RO}
\end{figure}

\begin{figure}[!h]
\centering
\makebox[\textwidth][c]{\includegraphics[width=0.7\textwidth]{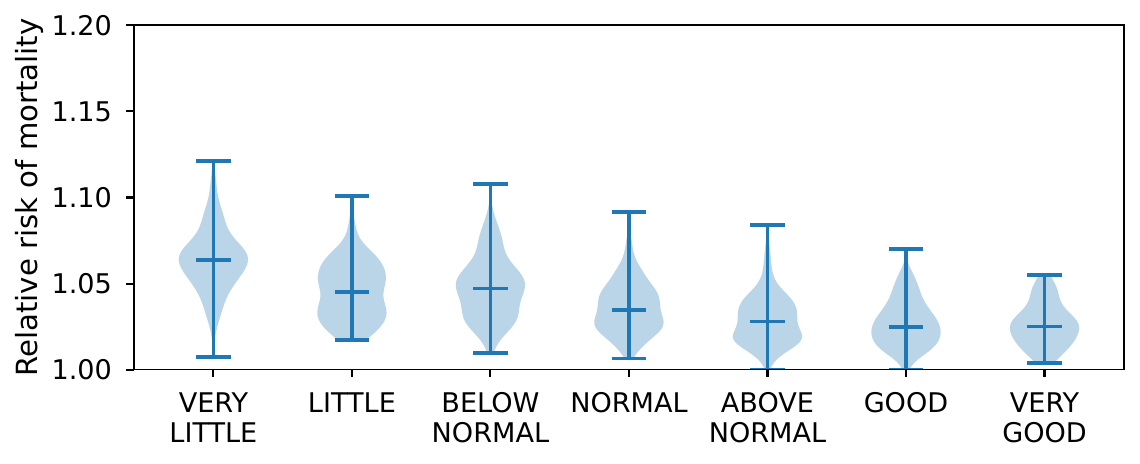}}
\vspace{-0.5 cm}
\caption{\textcolor{black}{Relative risk of mortality by building insulation level of residential buildings in `RO-HI' case}}
\label{fig:building_thermal_mortality_RO}
\end{figure}

\textcolor{black}{To provide further insights into how improving building insulation can influence the outcome, Figure~\ref{fig:building_thermal_RO}--\ref{fig:building_thermal_mortality_RO} illustrate the average temperature and the relative risk of mortality based on the insulation levels (thermal integrity parameters given in \cite{gld_house}) of residential buildings in the `RO-HI' case. We can observe an increasing trend in the mean value of average temperature as the building insulation level increases. The mean value of the average temperature in poorly insulated buildings is 12.4$^\circ$C, whereas, in well-insulated buildings, it is 17.4$^\circ$C, showing a 40\% increase. 
Accordingly, there is a noticeable increasing trend in the mean mortality rate as the building insulation level decreases. The relative risk increases from 1.026 for well-insulated buildings to 1.064 for those with minimal insulation. This finding highlights the crucial role of building insulation in improving indoor thermal conditions and mitigating the mortality risk during extreme events.}

\begin{figure}[t]
\centering
\makebox[\textwidth][c]{\includegraphics[width=1\textwidth]{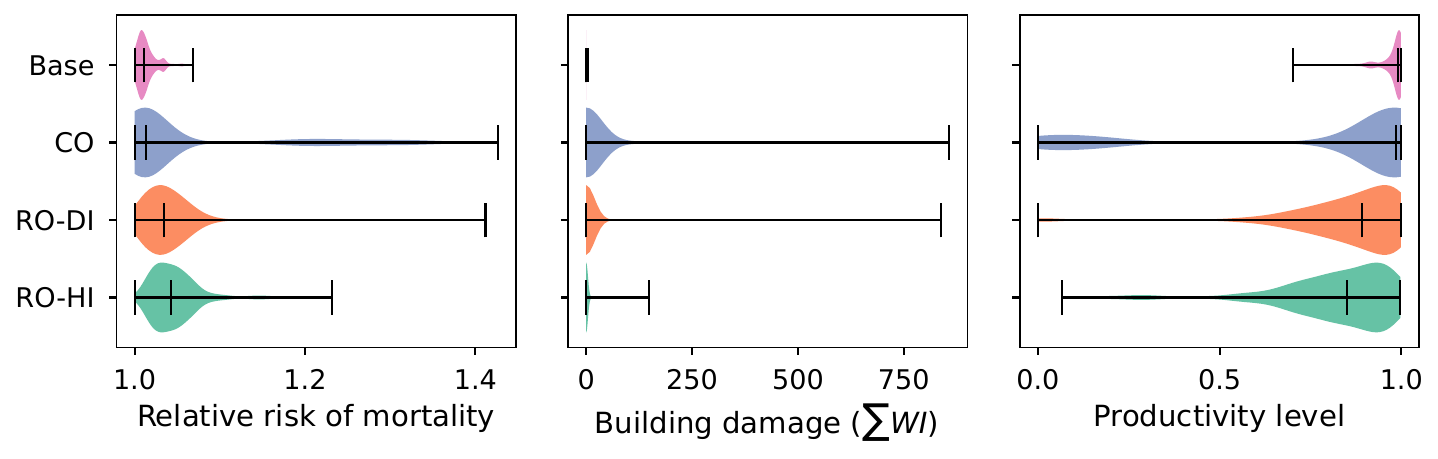}}
\vspace{-0.75 cm}
\caption{\textBlack{Relative risk of mortality, building damage ($\sum WI$), and productivity level for customers during the outage period (Feb $15^{th}$ - Feb $19^{th}$). Each plot compares the performances of the `CO', `RO-DI' and `RO-HI' cases with the `Base' case.}}
\vspace{-0.75 cm}
\label{fig:3plots}
\end{figure}

The exposure analysis has solidified our understanding that extremely cold weather during power outages can lead to substantial drops in indoor temperatures and potential damage to both occupants and buildings. To shed further light on this, Fig. \ref{fig:3plots} visualizes the loss levels in terms of the relative risk of mortality, building damage, and customer productivity, during the outage period, using violin plots.

In these violin plots, the minimum, maximum, and mean values are marked as black bars. Additionally, the width of each violin corresponds to the density of data points. \textcolor{black}{These measures help quantify and communicate the severity of the impact on human health and building damage during extreme outage events.} 
\textBlack{In  Fig. \ref{fig:3plots}, one observation is the variation in the relative risk of mortality across different cases. The maximum risk is increased up to 1.43 and 1.41 for the `CO' case and `RO-DI' case, respectively, while the `Base' case and `RO-HI' case maintain relatively lower risks below 1.06 and 1.23, respectively. 
It indicates that the mortality risk increases by 16 - 34\% during the extreme cold temperature-related outage.
Even within the same case, the mortality rate among customers varies depending on their building characteristics and resistance levels. The `CO' and `RO-DI' cases had the largest ranges indicating the occupants' health struggles against the extremely cold indoor temperature due to the outage.
When considering building damage levels, a substantial number of houses in the `CO' and `RO-DI' cases experience failing thermal conditions below critical values during outages, compared to the other two cases.}
As for productivity levels, the `CO' case has a notable density at values close to 1 and 0. This pattern is because some buildings without power contributed to a productivity level of 0, while others with power could maintain thermal conditions, leading to a higher productivity level close to 1. This finding demonstrates a significant inequity in thermal conditions by customers with the `CO' mechanism. 
\textBlack{
In contrast, the `RO-DI' and `RO-HI' cases, characterized by intermittent power supply, do not show clear distinctions among customers experiencing either good or bad thermal conditions during the outage event. Instead, it had lower productivity levels, corresponding to the relatively lower indoor temperatures during the outage.}
These observations provide insights into how building characteristics, individual resilience, and power availability play in determining the impact of cold-weather outages on mortality, building damage, and productivity.

\begin{figure}[ht]
\centering
\includegraphics[width=0.9\textwidth]{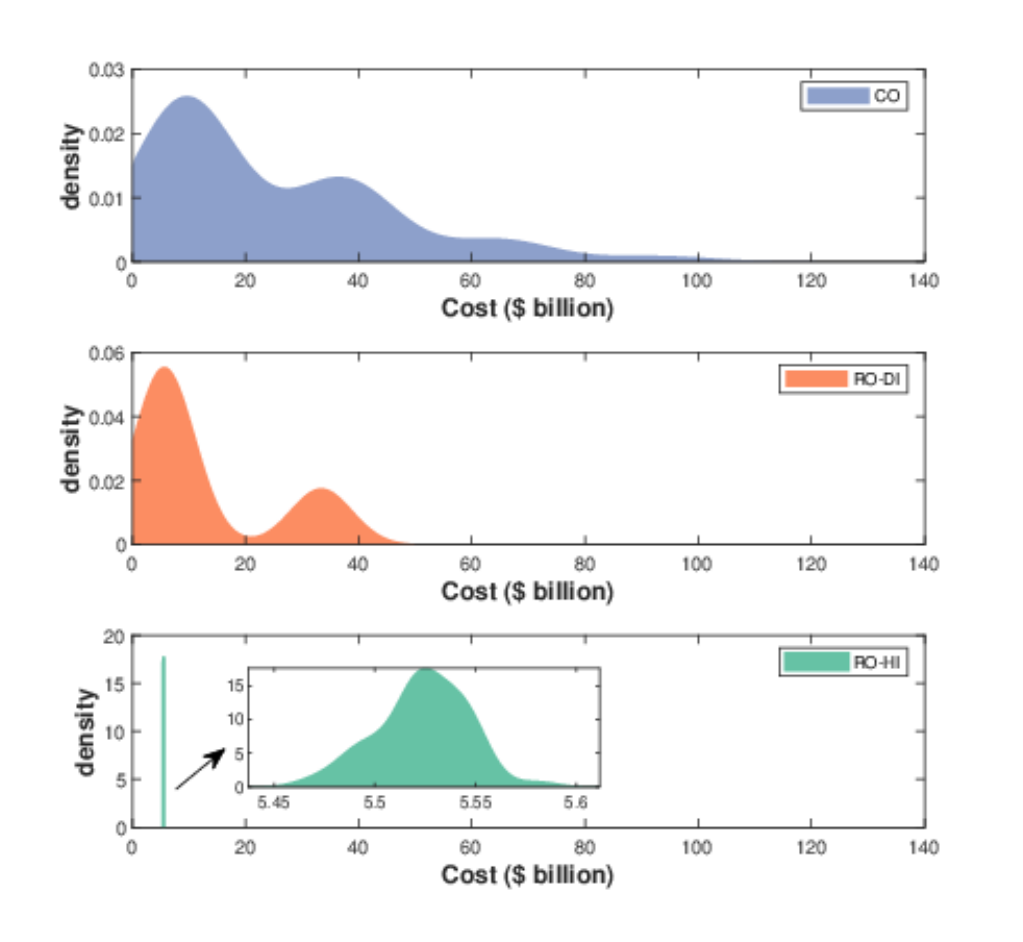}
 \vspace{-0.35 cm}
\caption{\textBlack{Costs distribution of `CO', `RO-DI', and `RO-HI' cases from Monte Carlo simulation}}
\label{fig:kde}
 \vspace{-0.25 cm}
\end{figure}

\begin{figure}[ht]
\centering
\includegraphics[width=0.85\textwidth]{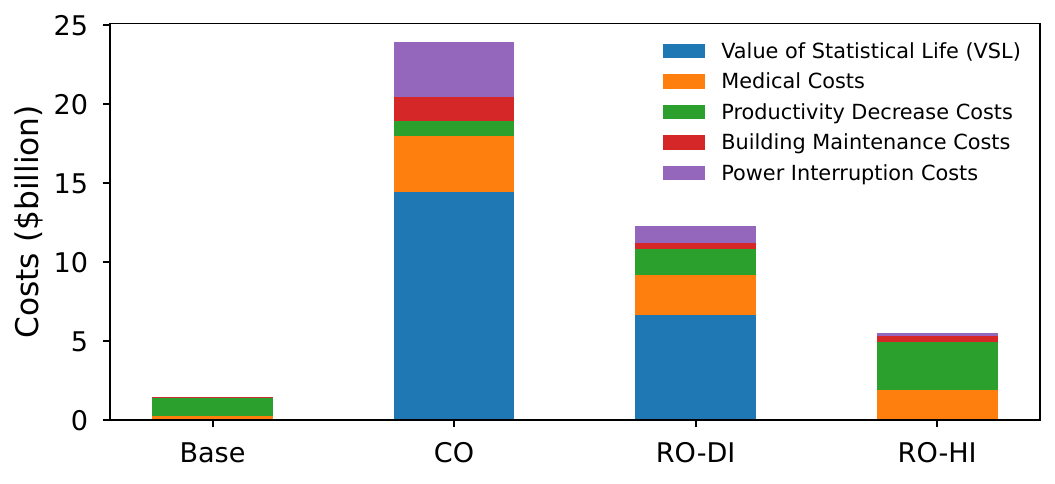}
 \vspace{-0.25 cm}
\caption{\textBlack{Comparison of stacked cost breakdown based on the mean value of cost distribution from Monte Carlo simulation}}
\label{fig:total_cost_barchart}
 \vspace{-0.25 cm}
\end{figure}

\subsubsection{Valuation}

\textBlack{Fig. \ref{fig:kde} provides a customer valuation of losses incurred during the outage, with the cost distributions of the `CO', 'RO-DI', and `RO-HI' cases derived from the Monte Carlo simulation. The plot is generated using the Kernel Density Estimate (KDE) to gain insights into the characteristics of the total costs in both cases. The simulated `CO' estimates a mean total damage of \$23.9 billion for the NC-WZ of Texas which is right in the ball-park of the estimated damage that was reported due to power outages during Winter Storm Uri \cite{FED}. This establishes a baseline for planning and evaluating the value of implementing resilience enhancement strategies. Furthermore, in the `CO' case, we observe a wider variation in total costs and longer tails, indicating the potential fluctuations in overall expenses and the possibility of extreme outlier costs. As for the `RO-DI' case, although it shows a narrower cost variation compared to the `CO' case, its cost distribution remains considerably wider and more visible compared to the `RO-HI' case, which shows the narrowest range in total costs. In the `RO-DI' case affected by damaged infrastructure, a mean total cost of \$12.3 billion which is almost half of the total cost observed in the `CO' case. Assuming no infrastructure damage, the potential of the enhanced strategy `RO-HI' was maximized by a mean total cost of \$5.52 billion, nearly a quarter of the total cost of \$23.9 billion for the `CO' case. 
This finding highlights the potential cost-effectiveness of the enhanced outage strategy `RO' in comparison to the `CO' strategy, due to more efficient resource allocation during scarcity. \\
To gain a deeper understanding of the breakdown of total costs, Fig. \ref{fig:total_cost_barchart} shows the comparison of stacked cost breakdowns across different cases. These cost breakdowns are based on the mean values derived from Monte Carlo simulations.
The main observation is the significant VSL costs associated with increased risks of mortality and medical costs associated with the `CO' and `RO-DI' cases. This is primarily due to a large number of occupants being exposed to extremely cold indoor temperatures even inside the buildings during the event, which can cause severe health issues such as hypothermia, heart stroke, and respiratory problems. Recent research has also included a measure of habitability, where a residence may no longer be considered habitable due to prolonged exposure to extreme indoor air temperatures, the cost of substitute housing was not included in this analysis but could be considered in future studies \cite{BTOPNNL}. Consequently, these health-related costs are reflected in the VSL and medical costs compared to the other cases. 
When comparing the `RO-DI' and `RO-HI' cases, the results indicate that even the enhanced outage scheme cannot fully demonstrate its potential due to external factors such as damaged infrastructure (as 3.4\% of customers were isolated due to faults for the `RO-DI' case), resulting in higher VSL and health-related costs as shown in the `RO-DI' case. Furthermore, the `CO' case has the highest building maintenance costs due to the substantial impact of extreme cold temperatures on building infrastructure. 
Meanwhile, in the `RO-HI' case, we observe the highest productivity decrease costs. This is because, with this outage mechanism along with hardened infrastructure, all the customers experienced intermittent outages, unlike the `CO' and `RO-DI' cases where a certain set of customers were not serviced at all throughout the outage period. 
Also, the indoor temperature conditions for the `RO-HI' case, are not optimal for higher productivity leading to increased loss due to productivity decrease. Inclusion of the opportunity cost of productivity losses should be assessed by the analyst when valuing a resilience solution to ensure that it is an appropriate cost to include given the population, customer type, and nature of the event.  }

Overall, we demonstrate the potential economic impact of extreme cold weather-related outages on customers and highlight the importance of incorporating detailed valuation into resilience planning and operations.
The results suggest that along with infrastructure upgrades, advanced outage management strategies have a huge impact on the total customer interruption costs, as they can be dramatically more expensive than normal operating conditions due to the cost of greater mortality risks and medical costs by extreme temperature-related outages. Therefore, it is crucial to consider these potential costs in investment decision-making while planning for enhancements in outage management systems. With this proposed framework, we can better assess the financial implications of each scenario, and help users make decisions regarding risk management and mitigation strategies.

\section{Discussion}
In this section, we highlight the significance and benefits of the established framework, encompassing the grid response model, simulated scenarios, and valuation showcase for both researchers and stakeholders. Additionally, we elaborate on the importance of this framework in planning in response to extreme events, emphasizing that investments must align with tangible results. 

\subsection{Significance of Simulation Testbed and Scenarios}
The proposed valuation framework is developed leveraging a modeling and simulation platform (TESP), that enables co-simulation of multiple simulation tools to emulate at-scale distribution grid response. Within the platform, GridLAB-D is used as a distribution system simulator, which in addition to providing traditional functionality, such as modeling power delivery elements, equipment loading, power flow, and voltage regulation also facilitates high-fidelity modeling of end-user loads through a mixture of ZIP (constant impedance, current, and power) loads driven by schedules and thermostatic loads models including water heaters and HVAC systems inside a building’s thermal envelope.  As described in section \ref{sec:grid_modeling}, the grid response model is developed based on regional characteristics, calibrated/fine-tuned against historical data, and then used to extract grid response under various hypothetical scenarios including business-as-usual (i.e., base case) and advanced outage management schemes. Hence, this simulation framework facilitates emulating the grid response considering the state of the grid infrastructure, load composition, and external variables such as extreme weather-induced demand spikes. The inclusion of weather, thermal dynamics of customers' end-uses, and power flow simulator make this platform well-suited for evaluating customer-oriented impacts of extreme temperature-related outages. To simulate outage scenarios, the ``meter'' attribute of a house object in GridLAB-D is leveraged, enabling the remote connect/disconnect of specific customer groups, and thereby providing the capability to simulate the impact of resilience enhancement strategies in response to extreme events.

\subsection{Innovative Customer-Oriented Valuation}
The overall approach of the valuation framework combines various elements such as temperature functions, non-energy impacts, probabilistic scenarios, customer-oriented factors, and grid simulations, which are beyond the existing traditional valuation methods. The customer characteristics, temperature functions \& probabilistic loss scenarios in the valuation framework capture the interruption costs and the non-energy impacts due to the dynamic nature of extreme temperature-related outages. This approach allows for a more realistic assessment of damage and loss, considering various scenarios and severity. Additionally, incorporating non-energy impacts (e.g., mortality) associated with extreme temperature-related outages provides a more thorough evaluation of the financial and societal consequences so that informed decision-making is available to enhance overall resilience and public safety. 

\subsection{Practical Implications}
The increasing frequency and intensity of extreme events leading to outages with catastrophic consequences has motivated grid planners and utility commissions to prioritize grid resilience and underscore the relationship and tradeoffs between resilience, and equity issues \cite{REG}.  However, some of the key technical challenges towards administering and justifying grid resilience-related investments can be attributed to the lack of analytical approaches that can translate grid performance into consequences for local communities and economies \cite{REG}. The proposed valuation framework presents a holistic assessment approach by integrating grid resilience enhancement measures (like outage-management automation or infrastructure hardening) with their impacts on reducing the frequency and severity of outages and the associated economic and social consequences. The proposed framework can guide resilience investment decisions by providing information about how alternative investments may improve grid resilience and the implications for local economies and vulnerable communities. Further, the disaggregation of performance-based metrics can also evaluate how the resilience benefits are distributed among the customers, thereby providing insights into the equity implications of alternative resilience investment strategies.

\section{Conclusion}

The valuation framework developed in this paper provides a valuable tool for understanding and estimating the customer-oriented impacts of extreme temperature-related outages. The integration of direct economic losses and non-energy impacts, such as mortality, into the damage and loss models provides a holistic perspective that goes beyond the existing resilience assessments. The damage/loss models with the function of temperature and probabilistic loss scenarios leveraging customer-oriented factors provide a different dimension to outage valuation. Incorporating the proposed resilience valuation with the scalable grid simulation environment further refines the ability of our approach to consider the diverse interplay of customer characteristics and grid response variables. 
\textBlack{This innovative approach takes into account the dynamic nature of extreme temperature-related outages and their effects on the communities.}

The real-world application of the proposed valuation framework is demonstrated using the 2021 winter storm Uri. The results showed that indoor temperatures can drop rapidly by -10°C during extreme temperature-related outages, highlighting the crucial role of electricity in maintaining thermal comfort for occupants. \textBlack{Exposure to these cold temperatures increased the relative risk of mortality by up to 34\%. }The findings align well with the report on Winter Storm Uri that 65\% of deaths due to hypothermia occurred inside homes without power \cite{TexasNews}. This underscores the importance of power availability and enhancing outage planning. \textBlack{Furthermore, the simulated `Controlled Outage (CO)' predicts an average total damage of \$23.9 billion for the NC-WZ of Texas, aligning closely with the estimated damage reported during Winter Storm Uri's power outages. Moreover, results indicate that just implementing an advanced outage management scheme (RO-DI), without any infrastructure upgrades, can reduce the relative risk of mortality by 1.4\%  and decrease the total costs related to non-energy impacts (excluding the cost of power interruption) by 45\%. Finally, we demonstrated that the enhanced outage mechanism with improved infrastructure can mitigate non-energy impacts by reducing the relative risk of mortality by 16\% and saving total costs related to non-energy impacts by 74\%. This establishes a foundational benchmark for assessing the effectiveness of resilience enhancement strategies and planning future strategies.}

The presented valuation demonstrated specifically for outages caused by extremely cold temperatures, which would require adjustment to be applicable for extreme heat scenarios (e.g. see \cite{Astrom2011, Anderson2011}). 
The proposed approach is limited by several assumptions (e.g., transportation availability, socioeconomic status) and may provide rough estimates as it utilizes the variables from simulations. 
In addition, this study did not consider the potential damage to grid infrastructure caused by the extreme weather. Future research can improve these limitations. However, the framework's capability of generating resilience scenarios based on a grid simulation environment can be utilized by researchers, analysts, and planners to quantify the cost and non-energy impact of grid outages. This valuation framework offers a more comprehensive understanding of the economic impact of power outages, enabling well-informed decisions regarding investments in resilience planning. Further research is required with more collaboration among public health experts and other agencies. As we face an era of increasing climate variability, this framework stands as a vital tool for policymakers, utility planners, and researchers striving to build more resilient and sustainable energy infrastructures.

 \bibliographystyle{elsarticle-num} 
 \bibliography{main}

\newpage
\appendix

\section{Valuation Framework Parameters} \label{appendix}
Appendix A offers a comprehensive overview of the fundamental parameters and variables employed in our valuation framework, particularly for demonstration purposes. 
Note that these parameters are presented as samples for our demonstration work. We strongly recommend tailoring the parameters to suit your specific case. This work primarily serves as a guide to understanding the valuation framework rather than prescribing specific parameters. 

Key statistics for damage function models and financial analysis of our study are summarized in Table \ref{tab:statistics_function_models}.

\begin{table}[h]
\centering
\caption{Statistical parameters}
\label{tab:statistics_function_models}
 \begin{adjustbox}{max width=\textwidth}
\begin{tabular}{c|c|c|c|c|c}
\hline
\multicolumn{2}{c|}{} & mean & std & min & max \\
\hline
Pre-existing &Cardiac & 5.1 & 1.0 & 0 & 100 \\
\cline{2-6}
health conditions (\%) & Respiratory & 7.3 & 1.0 & 0 & 100 \\
\hline
\multicolumn{2}{c|}{Health insurance (\%)} & 79.4 & 3.0 & 0 & 100 \\ 
\hline
\multicolumn{2}{c|}{Healthcare accessibility (\%)} & 89.4 & 3.0 & 0 & 100 \\
\hline
\multirow{3}{*}{Hospital survivability (\%)} &Cardiac & 89.3 & 1.0 & 0 & 100 \\
\cline{2-6}
& Respiratory & 83.0 & 1.0 & 0 & 100 \\
\cline{2-6}
& Hypothermia/Frost & 91.9 & 3.0 & 0 & 100 \\  
\hline
\multirow{3}{*}{Home survivability (\%)} &Cardiac & 19.3 & 1.0 & 0 & 100 \\
\cline{2-6}
& Respiratory & 13.0 & 1.0 & 0 & 100 \\
\cline{2-6}
& Hypothermia/Frost & 78.9 & 1.0 & 0 & 100 \\ 
\hline
\multirow{3}{*}{Medical cost w/insurance (\$)} & Cardiac & & & 1014 & 6282 \\
\cline{2-6}
&Respiratory  &&&1014 & 6282 \\
\cline{2-6}
&Hypothermia/Frost  &&&1014 & 6282 \\ 
\hline
\multirow{3}{*}{Medical cost w/o insurance (\$)} & Cardiac & & & 3162 & 15348 \\
\cline{2-6}
&Respiratory  &&&3162 & 15348 \\
\cline{2-6}
&Hypothermia/Frost  &&&3162 & 15348 \\
\hline
\multicolumn{2}{c|}{Home insurance (\%)} & 95.9 & 3.0 & 0 & 100 \\  
\hline
\multicolumn{2}{c|}{Pipe repair cost w/insurance (\$)}   & & & 500 & 2000 \\
\hline
\multicolumn{2}{c|}{Pipe repair cost w/o insurance (\$)}   & & & 600 & 5000 \\
\hline
\toprule[0.5 mm]
\end{tabular}
\end{adjustbox}
\end{table}

\begin{table}[h]
\centering
\caption{Average hourly pay rate based on building type}
\label{tab:average_wage}
\begin{tabular}{l|l}
\hline
Building Type & Average hourly pay rate (\$) \\
\hline
single family & 45.51 \\
\hline
multi-family  & 45.51 \\
\hline
mobile home   & 45.51 \\
\hline
office & 37.88 \\
\hline
warehouse storage & 15.49 \\
\hline
big box & 29.36 \\
\hline
strip mall & 22.38 \\
\hline
education & 27.95 \\
\hline
food service & 13.4 \\ 
\hline
food sales & 15.64 \\
\hline
lodging & 13.44 \\
\hline
healthcare & 43.15 \\
\hline
low occupancy & 21.41 \\
\toprule[0.5 mm]
\end{tabular}
\end{table}

We use the percentages of residents with cardiovascular and respiratory disease in Texas \cite{disease_percent} for pre-existing health conditions. As for health insurance rates, we refer to the values from the Texas Medical Association \cite{texas_medical}.
Regarding healthcare accessibility, we originally used the health insurance rate because lack of health insurance coverage can lead to reduced access to healthcare services, and then adjusted considering other factors that might lower the accessibility during cold temperature events. 
For improved chances of survival, individuals can benefit greatly from timely access to healthcare facilities. Especially, this is crucial for those with pre-existing health conditions, as their survivability rates significantly increase once they reach an emergency room, hospital, or urgent care center, given their need for extensive medical equipment. 
Even for individuals without pre-existing health problems, survivability rates are comparatively higher when healthcare access is available. The values mentioned here are based on data from various sources \cite{Ko2019, Ranasinghe2023, Stefan2013}. However, these values are rough estimates and have been adjusted for the purposes of this study.
Medical costs are based on Anthem's `Find Care \& Estimate Costs' feature available on the platform \cite{anthem}. 
As for home insurance rate, we refer to the values from Texas Real Estate Research Center \cite{texas_realestate_center}. The productivity decrease cost, which represents the economic impact of potentially reduced productivity due to discomfort caused by indoor conditions, is estimated based on the number of workers in building along with the average hourly wage associated of the job categories. Table \ref{tab:average_wage} 
presents the a list of the building types considered in this work along with the  average hourly wage of the associated jobs for Texas that were obtainted from U.S. Bureau of Labor Statistics \cite{blsTX}. Building maintenance and repair costs consider frozen pipes. We refer to multiple resources to get the values. To get an accurate estimate for pipe repair, it's best to contact a licensed plumber or pipe repair specialist who can assess the specific situation and provide a detailed cost estimate.

\end{document}